\documentclass[]{raa}
\usepackage[noblocks]{authblk}
\usepackage{cases}            
\usepackage{graphicx,times}
\usepackage{url}
\usepackage{pdflscape}
\usepackage{longtable}
\usepackage{multirow}

\begin{document}

\title{Spectral Energy Distributions for TeV Blazars}

\volnopage{ {\bf 201X} Vol.\ {\bf XX} No. {\bf XX}, 000--000}
\setcounter{page}{1}

\author{C. Lin
  \inst{1,2} \footnote{\it linchao@e.gzhu.edu.cn}
\and J. H. Fan
  \inst{1,2} \footnote{Corresponding author: {\it fjh@gzhu.edu.cn}}
  }

\institute{Center for Astrophysics, Guangzhou University,  Guangzhou 510006, China;
    \and
    Astronomy Science and Technology Research Laboratory of Department of Education of Guangdong Province, Guangzhou 510006, China \vs \no \\
    {\small Received [year] [month] [day]; accepted [year] [month] [day]}
}

\abstract{
In this work, we collected a sample of 69 TeV blazars from TeVCat,
obtained their multi-wavelength observations,
and fitted their spectral energy distributions by using the second degree polynomial function.
The structure parameters of the synchrotron bumps for 68 blazars and
those of the inverse-Compton bumps for 56 blzars are obtained.
Then, we adopted statistical analysis to the parameters (
 curvature,
 peak frequency,
 peak luminosity,
 bolometric luminosity, and
 X/$\gamma$-ray spectral indexes).
 From our analysis and discussions,
   we can get following conclusions:
1. There is a clear positive correlation between
    the synchrotron peak frequency, $\log \nu_{\rm p}^{\rm s}$, and
    the inverse-Compton peak frequency $\log \nu_{\rm p}^{\rm IC}$, and that
    between the synchrotron peak luminosity, $\log \nu_{\rm p}^{\rm s}L_{\nu_p}^{\rm s}$, and
    the inverse-Compton peak luminosity, $\log \nu_{\rm p}^{\rm IC}L_{\nu_p}^{\rm IC}$.
2.  The correlation between the peak frequency and the curvature of synchrotron bump
    is clearly different from that of the inverse-Compton bump,
    which further indicates that there are different emission mechanisms between them.
3.  There is a correlation between $\log \nu_{\rm p}^{\rm IC}$ and $\gamma$-ray spectral index,
    $\alpha_{\gamma}$, for the TeV blazars:
    $\log \nu^{IC}_{p} = -(4.59 \pm 0.30) \alpha_{\gamma} + (32.67 \pm 0.59)$,
    which is consistent with previous work of Abdo et al.(2010).
4. An "L-shape" relation is found between $\log \nu^{s}_{p}$ and $\alpha_X$ for both TeV blazars and Fermi blazars.
    A significant correlation between $\log \nu^{s}_{p}$ and X-ray photon index ($\alpha_X$)
     is found for the TeV blazars with high synchrotron peak frequency:
     $\log \nu^{s}_{p} = -(3.20 \pm 0.34) \alpha_X + (24.33 \pm 0.79)$, while the correlation is positive for low synchrotron peaked TeV sources.
5.  In the $\alpha_X-\alpha_\gamma$ diagram, there is  also an "L-shape", the anti-correlation is consistent with the
available results in the literature, we also find a positive correlation between them.
6. Inverse-Compton dominant sources have luminous bolometric luminosities.
}

\authorrunning{C. Lin \& J. H. Fan}            
\titlerunning{Spectral Energy Distributions for TeV Blazars}  
   \maketitle

\section{Introduction}

Blazars are a subclass of the radio-loud active galactic nuclei (AGNs) with extreme observational properties,
  such as high and variable polarization,
  rapid and large variability in the  non-thermal continuum,
  strong $\gamma$-ray emissions, and
  superluminal motions, etc
  (Urry \& Padovani 1995;
  Fan 2005).
In a standard AGN  model,
  there is a supermassive black hole surrounded by an accretion disk at the center,
  and there is a high speed jet perpendicular to the accretion disk.
Blazars are believed to be a subclass of AGNs whose jet axis is very close to the line of the observer's sight.
Blazars are divided into flat spectrum radio quasars (FSRQs) and BL Lacertae objects (BL Lacs).
BL Lacs and FSRQs show similar continuum emission properties with
  BL Lacs showing weak (or even no) emission line features but
  FSRQs displaying strong emission lines
(Aller et al. 1992;
 Urry \& Padovani 1995;
 Fan et al. 2009;
 Hovatta et al. 2009;
 Lin et al. 2017).

Recently,
  the emissions of TeV band,
  which are also called as very-high-energy band (VHE; $E>100$ GeV),
  are detected from an increasing number of extragalactic sources (Acero et al. 2015).
TeVCat\footnote{\url{http://tevcat.uchicago.edu/}} is an online catalog for VHE $\gamma$-Ray Astronomy,
  collecting the sources that have been detected by the ground-base telescopes (Wakely \& Horan 2008).
TeV emissions are important to constrain the emission mechanism of AGNs and studied by some researchers
(eg.,
  Weekes 1997;
  Giannios et al. 2009;
  Piner et al. 2010;
  Holder  2012;
  Zhang et al. 2012;
  Xiong et al. 2013;
  Aartsen et al. 2015;
  Lin \& Fan 2016;
  Sahu et al. 2016).

The broad-band SEDs of blazars show a double bump structure in the flux ($\log \nu f_{\nu}$) versus frequency ($\log \nu$) panel.
The low energy bump commonly peaks at infrared to X-ray bands,
  while the high energy one peaks at MeV-GeV band
 (Fossati et al. 1998;
  Kang et al. 2014).
It is generally accepted that the low energy bump is corresponding to a synchrotron emission
  which is produced by the relativistic electrons in the jet,
  while the high energy bump is corresponding to an inverse-Compton (IC) emission
  which is caused by the IC scattering.
However, different authors have different ideas about the origins of the seed photons of IC process.
Some researchers think that IC process is from synchrotron self-Compton (SSC) process with the seed photons
 being from synchrotron emissions in the jet
(eg., Rees et al. 1967;
      Jones et al. 1974;
      Marscher \& Gear 1985;
      Mastichiadis et al. 1997;
      Krawczynski et al. 2001;
      Sikora et al. 2001;
      Abdo et al. 2014;
      Hovatta et al. 2015;
      Yang et al. 2017a,b).
Others believe that the seed photons are from exterior of the jet,
  such as accretion disk, broad line region,
  dust torus and so on,
  and that is called external Compton (EC) process
(eg., Dermer et al. 1992;
      Dermer \& Schlickeiser 1993;
      Sikora et al. 1994;
     Hartman et al. 2001;
     Sokolov \& Marscher 2005;
     Albert et al. 2008).
Another classification of blazars is the  one based on their
 synchrotron peak frequency of the spectral energy distributions (SEDs)
 (Padovani \& Giommi 1995;
  Nieppola et al. 2006;
  Abdo et al. 2010a;
  Ackermann et al. 2015;
  Lin \& Fan 2016;
  Fan et al. 2016
  ).
In our previous work (Fan et al. 2016), we divided blazars into
low synchrotron peaked (LSP, $\nu_{\rm{peak}}^{s} <10^{14}$ Hz),
intermediate synchrotron peaked  (ISP, $10^{14}$ Hz $<\nu_{\rm{peak}}^{s} < 10^{15.3}$ Hz), and
high synchrotron peaked (HSP, $\nu_{\rm{peak}}^{s} > 10^{15.3}$ Hz) blazars by analyzing a sample of 1329 blazars.

In this work, we collect a sample of 69 TeV blazars from TeVCat,
     and try to keep track of all announced TeV sources,
     so that our sample is the largest for TeV blazars until April 2018.
Because only one TeV blazars (1943 + 213) is not detected by Fermi telescope,
    and the dominant power of IC bump is at Fermi observational band,
    we assume that TeV blazars are, not strictly, a sub-sample of Fermi blazars only for statistical comparison in this work.
We use a second degree polynomial function to fit their multiwavelength SEDs,
     obtain fitting parameters,
     calculate monochromatic luminosity and bolometric luminosity.
This paper is arranged as follows:
We will give a sample and fitting procedure in Section 2,
 results in Section 3,
 some discussions and conclusions in Sections 4 and 5.

\section{Sample and Fitting Procedure}

From TeVCat,
    we found that there are 210 sources detected in TeV energy range until April 2018,
    and 75 (69 blazars) of them are the extragalactic TeV sources with low redshift ($z<1.0$).
Most of the TeV blazars are HSP BL Lacs,
    which are usually considered to be the candidates of the extragalactic TeV sources
(Lin \& Fan 2016, and reference therein).
The connection between TeV source and peak frequency is still an interesting topic of blazars.

A useful tool for building SEDs in Space Science Data Center (SSDC) is
    SSDC SED Builder\footnote{\url{http://tools.asdc.asi.it/SED/}},
    which combines radio to $\gamma$-ray (even TeV) band data from several missions and
    experiments together with catalogs and archival data
(Aharonian et al. 2001;
Giommi et al. 2002;
Amenomori et al. 2003;
Aharonian et al. 2003, 2009;
Daniel et al. 2005;
Schroedter et al. 2005;
Acciari et al. 2008, 2011;
Godambe et al. 2008;
Chandra et al. 2010, 2012;
H.E.S.S. Collaboration 2010, 2013;
Aliu et al. 2011, 2015;
Bartoli et al. 2011, 2012;
Archambault et al. 2013, 2014;
Arlen et al. 2013;
Abramowski et al. 2013, 2015;
Biteau \& Williams 2015;
Sharma et al. 2015).
In this work, we collect 69 blazars (62 BL Lacs and 7 FSRQs) from TeVCat, and obtain their SEDs by using SSDC SED Builder.
We use the default search radius values in SSDC SED Builder to search the observational data from the corresponding databases,
    and reject the data whose flux error is larger than flux and the upper limit data.
We find that some observational data ($\nu >10^{16}$ Hz) from the NASA/IPAC Extragalactic Database (NED)\footnote{\url{http://ned.ipac.caltech.edu/ }} is significantly different from other databases. Because the data from NED is collected from different archives which use different methods to calculate the observational data, thus we neglected the data whose $\nu >10^{16}$ Hz from NED in this work.

In order to unify the fitting procedure for each source,
    we did not consider the observational time of the data,
    unless the wide range of time causing the fitting result unreliable.
For example, if several times of flare are detected in TeV band,
we choose an time interval to keep the latest observation of flare.
Observations indicate that the thermal radiation is exist and causes a bump at UV/optical bands of SEDs for some AGNs
  (eg.,
  Shields 1978;
  Czerny \& Elvis 1987;
  Ross et al. 1992;
  Mannheim et al. 1995;
  Pounds et al. 1995).
We exclude the thermal radiation of the source by visual inspection,
    because the thermal radiation is significantly greater than
    the non-thermal radiation at UV/optial bands.
Similarly, we determine the break point between two bumps by visual inspection,
    and then fit the two bumps respectively.

In this work, we found that the SEDs of some sources are asymmetrical in synchrotron bump, such as 0721+713, especially in low energy radio band ($\log \nu < 10^9$ Hz).
Since the second degree polynomial function is symmetrical, it is better that those SEDs are fitting by a third degree polynomial function.
The third degree polynomial function is an empirical fitting function in SEDs calculations, and used in many works (eg., Fossati et al. 1998; Kubo et al. 1998; Abdo et al. 2010b; Wu et al. 2011; Ackermann et al. 2015; Chang et al. 2017). The asymmetrical structures of SEDs are possibly produced by some physical process, such as the synchrotron self-absorption process.

But, there are also many works indicate that the structure of SEDs is symmetric for blazars (eg., Landau et al. 1986; Sambruna et al. 1996; Zhang et al. 2002; Massaro et al. 2004a, b, 2006; Nieppola al. 2006; Wu et al. 2009; Meyer et al. 2011; Xue et al. 2016; Fan et al. 2016).
Massaro (2004b) found that the log-parabolic spectral model is not only the simplest way to fit the SEDs of Mrk 421,  but also relates to the physics of the statistical acceleration process under some simple hypotheses.
In the work of Xue et al. (2016), they discussed the difference of luminosity between the second and third degree polynomial fitting for some sources with apparent asymmetrical synchrotron bump. They found that the logarithm of luminosity in third degree polynomial fitting is lower than that in second degree polynomial fitting, and the differences are within 1.

The three parameters of the second degree polynomial function can give the basic structure of the SEDs,
    namely, the spectral curvature,
    the peak frequency,
    and the peak flux.
For analyzing the basic structure of the SEDs for blazars,
    we prefer to use the second degree polynomial,
    the reasons are:
1. the second degree polynomial is enough for determining the basic structure;
2. Most of the SEDs of TeV blazars are symmetrical in this work;
3. As discussed in Xue et al. (2016), the difference of luminosity between the second and third degree polynomial fitting
is insignificant;
4. It is hard to give a meaningful definition of each parameter for the third degree polynomial.

Following our previous work (Fan et al. 2016),
    we can fit the synchrotron bump and IC bump of SEDs using the following function:
\begin{equation}
  \log (\nu f_{\nu}) = c (\log \nu - \log \nu_{\rm p})^2 + \log \nu_{\rm p}f_{\nu_{\rm p}},
\end{equation}
    where $|2c|$ is the spectral curvature,
    $\log \nu_{\rm p}$ is the logarithm of  peak frequency,
    and $\log \nu_{\rm p}f_{\nu_{\rm p}}$ is the logarithm of peak flux.
In order to reduce the effect of fitting from asymmetrical SEDs, we set the lower limit of the frequency to $\nu = 10^9$ Hz in the second degree polynomial fitting.
Comparing with other energetic bands,
    optical band has plenty observational data for some sources.
Consequently,
    when the least square method is adopted to fit the SEDs,
    the second degree polynomial fitting will be dominant by the optical band.
To rebalance the weights between different bands,
    we calculate the averaged fluxes and frequencies for each frequency bin by using the following formulas:
\begin{equation}
  \log (\nu f_{\nu}) = \sum^{n}_{i=1}\frac{\log (\nu_{i} f_{i})}{err_{i}^2}/\sum^{n}_{i=1}\frac{1}{err_{i}^2},
\end{equation}
\begin{equation}
  \log \nu = \sum^{n}_{i=1}\frac{\log \nu_{i}}{err_{i}^2}/\sum^{n}_{i=1}\frac{1}{err_{i}^2},
\end{equation}
where $err^2=err_{+}^2+err_{-}^2$,
      $err_{+}$ and $err_{-}$ are the positive and negative errors of $\log (\nu f_{\nu})$,
      and we set the bin width to be $\Delta (\log \nu) = 0.1$.
For the flux without error, we use the averaged value of neighboring errors to replace it.

Peak luminosity ($\nu_{p} L_{p}$) and bolometric luminosity ($L_{bol}$) in source rest frame are  calculated using the following formulas:
\begin{equation}
  \nu_{p} L_{p}=4 \pi d_L^2 \nu_{p}' f_{\nu_{p}}'
\end{equation}
\begin{equation}
L_{bol}
  =4 \pi d_L^2 \ln(10) \int_{\log\nu_L}^{\log\nu_U} 10^{c (\log \nu - \log \nu_{\rm p}')^2 + \log \nu_{\rm p}'f_{\nu_{\rm p}}'} {\rm d}(\log\nu)
\end{equation}
where
    $d_{\rm L} =
 (1+z)\frac{c}{H_{0}}\int^{1+z}_{1}\frac{1}{\sqrt{\Omega_{M}x^{3}+1-\Omega_{M}}} \,\,{\rm d}x$ is luminosity distance,
    $\nu_p' = (1 + z) \nu_p$,
    $f_{\nu_p}' = f_{\nu_p} (1+z)^{\alpha_p -1}$,
    $(1+z)^{\alpha_p-1}$ stands for K-correction,
 and $\alpha_p = - \partial \log f_{\nu} / \partial \log \nu = 1$ is a spectral index at the peak frequency.
 Average value is adopted in K-correction if the redshift of the source is not available.
In this work, we set the lower and upper limits of integration to be $\nu_L=10^9$Hz and $\nu_U =10^{27}$Hz.
We adopt $\Omega_{\Lambda}\simeq0.7$,
    $\Omega_{\rm M}\simeq0.3$ and $\Omega_{\rm K}\simeq0.0$ from the $\Lambda$-CDM model (Capelo \& Natarajan 2007),
    and $H_0 = 67$ km s$^{-1}$ Mpc$^{-1}$.

\section{Results}

\begin{figure}
\centering
\includegraphics[width=6.5 in]{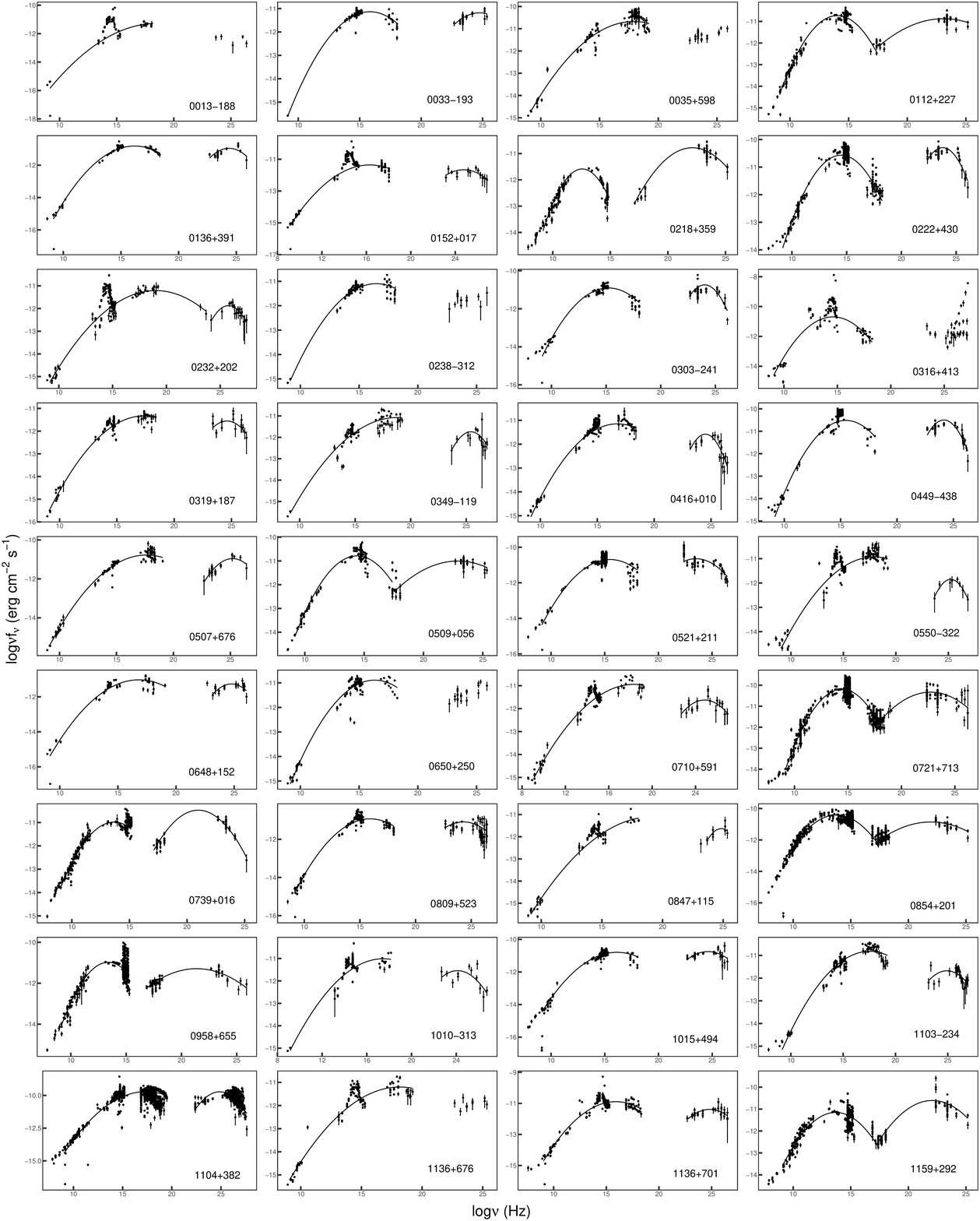}
\caption{\label{ms0075fig1} Spectral Energy Distributions for TeV Blazars.}
\end{figure}

\begin{figure}
\centering
\includegraphics[width=6.5 in]{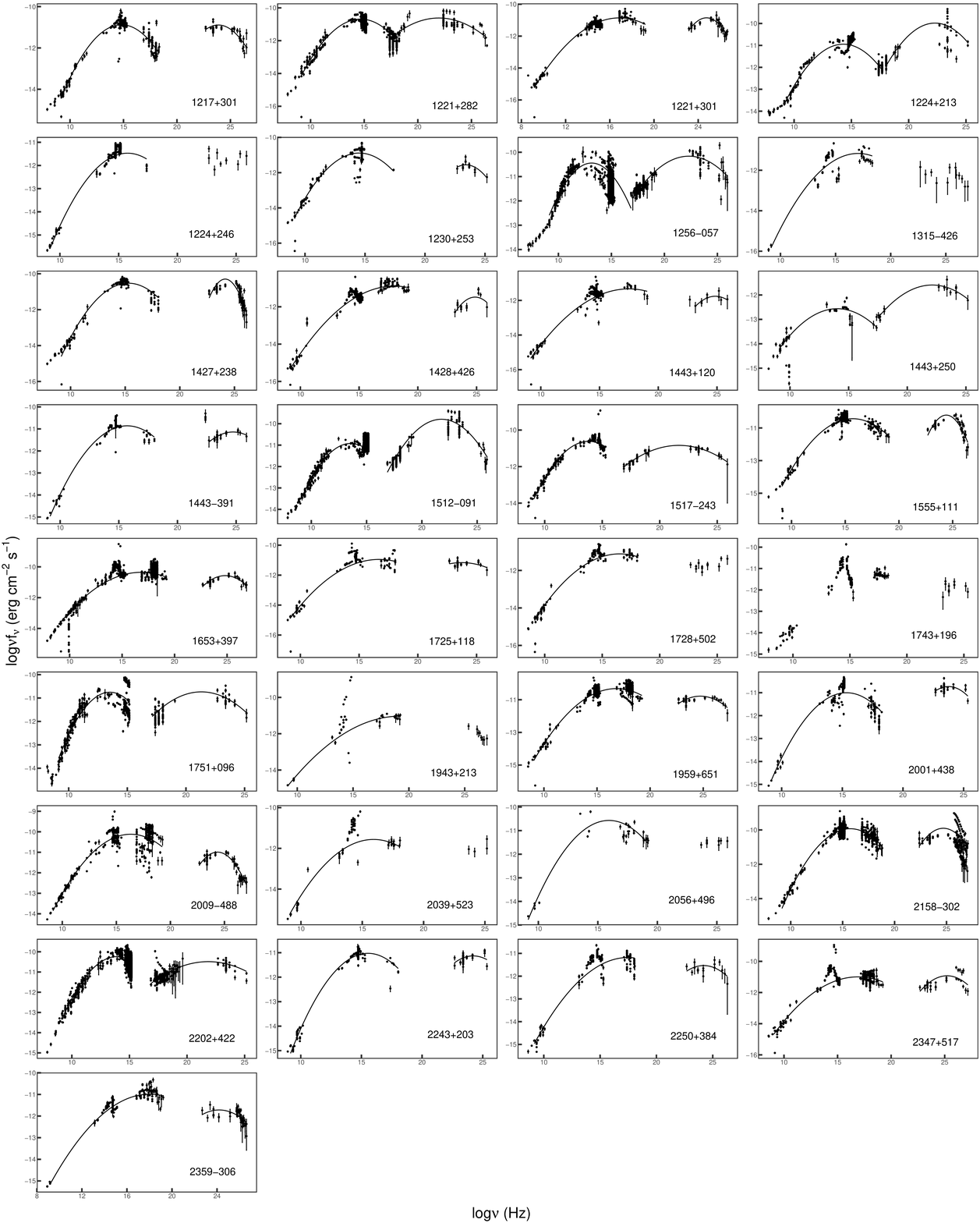}
\caption{\label{ms0075fig2} Spectral Energy Distributions for TeV Blazars.}
\end{figure}

We fit the SEDs for the whole 69 TeV blazars using the second degree polynomial function.
Because the observations are scarce or do not cover enough range of frequency,
    we can not fit the SEDs reliable for some sources.
Finally, the fitting results of synchrotron bump for 68 blazars and those of IC bump for 56 blzars are obtained.
The SEDs fitting results for the 69 blazars are shown in Figures \ref{ms0075fig1}-\ref{ms0075fig2}.
In Figures \ref{ms0075fig1}-\ref{ms0075fig2},
 the thermal radiation data are included in the figure but they are excluded in the SED fitting.
The curves in the figures stand for the fitting results.
If some time intervals are selected for some sources in TeV band,
then only the data in the chosen time interval are shown in the figures.

For statistical analysis,
    X-ray spectral index from Swift-XRT point source catalog
    (1SXPS, Evans et al. 2014) and $\gamma$-ray spectral index from the third Fermi Large Area Telescope source catalog (3FGL) are also collected.
The sample and fitting results in observer frame are shown in Table 1.
Particularly, the redshift effect of peak luminosity and bolometric luminosity are already corrected, see Equations (4) and (5).
In Table 1,
 Col. (1) gives TeVCat name;
 Col. (2) other name, the sources with "$^*$" are FSRQs, the rests are BL Lacs;
 Col. (3) redshift ($z$);
 Col. (4) the spectral curvature ($|2c^{s}|$) of synchrotron bump and its uncertainty;
 Col. (5) synchrotron peak frequency ($\log \nu^{s}_{p}$) in units of Hz and its uncertainty;
 Col. (6) synchrotron peak flux ($\log \nu^{s}_{\rm p}f_{\nu^{s}_{\rm p}}$) in units of erg/cm$^2$/s and its uncertainty;
 Col. (7) the spectral curvature ($|2c^{IC}|$) of IC bump and its uncertainty;
 Col. (8) IC peak frequency ($\log \nu^{IC}_{p}$) in units of Hz and its uncertainty;
 Col. (9) IC peak flux ($\log \nu^{IC}_{\rm p}f_{\nu^{IC}_{\rm p}}$) in units of erg/cm$^2$/s and its uncertainty;
 Col. (10) bolometric luminosity ($\log L_{bol}$) in units of erg/s;
 Col. (11) $\gamma$-ray photon index ($\alpha_{\gamma}$) from 3FGL;
 Col. (12) X-ray photon index ($\alpha_X$) from 1SXPS.
The statistical values of the fitting results are show in Table 2. In Table 2 and the following statistical analysis, frequencies, fluxes and luminosities are corrected to the sources' rest-frame.

\newpage
\begin{landscape}
\clearpage
\begin{center}
\begin{small}
\begin{longtable}{|cccccccccccc|}
 \caption{The Fitting Results in observed frame for 69 TeV Blazars.}\\
 \hline
TeVCat name &
other name &
$z$ &
$|2c^{s}|$/$\sigma$ &
$\log \nu^{s}_{p}$/$\sigma$ &
$\log \nu^{s}_{\rm p}f_{\nu^{s}_{\rm p}}$/$\sigma$ &
$|2c^{IC}|$/$\sigma$ &
$\log \nu^{IC}_{p}$/$\sigma$ &
$\log \nu^{IC}_{\rm p}f_{\nu^{IC}_{\rm p}}$/$\sigma$ &
$\log L_{bol}$ &
$\alpha_{\gamma}$ &
$\alpha_X$ \\\hline
 (1) & (2) & (3) & (4) & (5) & (6) & (7) & (8) & (9) & (10) & (11) & (12)\\
 \endfirsthead
 \caption{Continue.}\\
 \hline
TeVCat name &
other name &
$z$ &
$|2c^{s}|$/$\sigma$ &
$\log \nu^{s}_{p}$/$\sigma$ &
$\log \nu^{s}_{\rm p}f_{\nu^{s}_{\rm p}}$/$\sigma$ &
$|2c^{IC}|$/$\sigma$ &
$\log \nu^{IC}_{p}$/$\sigma$ &
$\log \nu^{IC}_{\rm p}f_{\nu^{IC}_{\rm p}}$/$\sigma$ &
$\log L_{bol}$ &
$\alpha_{\gamma}$ &
$\alpha_X$ \\\hline
 (1) & (2) & (3) & (4) & (5) & (6) & (7) & (8) & (9) & (10) & (11) & (12)\\
 \hline
\endhead
 \hline
\endfoot
 \hline
\endlastfoot
 \hline

J0013-188 & SHBLJ001355.9-1854 & 0.094 & 0.10/0.01  & 18.56/0.58  & -11.28/0.10  & & & &  & 1.94  & 2.12  \\
J0033-193 & KUV00311-1938 & 0.61 & 0.20/0.01  & 15.82/0.05  & -11.13/0.02  & 0.24/0.13  & 24.75/0.41  & -11.18/0.06  & 47.41  & 1.71  & 2.47  \\
J0035+598 & 1ES0033+595 & 0.086 & 0.11/0.01  & 17.91/0.40  & -10.68/0.06  &  &  &  &  & 1.90  & 2.04  \\
J0112+227 & S20109+22 & 0.265 & 0.30/0.01  & 14.03/0.07  & -10.73/0.05  & 0.08/0.01  & 23.11/0.32  & -10.88/0.04  & 46.87  & 2.03  & 1.87  \\
J0136+391 & RGBJ0136+391 &  & 0.18/0.01  & 16.18/0.07  & -10.81/0.02  & 0.43/0.21  & 24.37/0.20  & -10.96/0.15  &  & 1.70  & 2.41  \\
J0152+017 & RGBJ0152+017 & 0.08 & 0.15/0.01  & 16.32/0.18  & -11.36/0.04  & 0.27/0.10  & 24.75/0.19  & -11.66/0.10  & 44.96  & 1.89  & 2.51  \\
J0218+359 & S30218+35$^*$ & 0.944 & 0.41/0.03  & 12.56/0.05  & -11.59/0.07  & 0.17/0.01  & 22.13/0.13  & -10.78/0.07  & 48.14  & 2.28  & 1.75  \\
J0222+430 & 3C66A & 0.444 & 0.24/0.01  & 14.50/0.05  & -10.57/0.05  & 0.46/0.11  & 23.77/0.15  & -10.29/0.09  & 47.68  & 1.94  & 2.27  \\
J0232+202 & 1ES0229+200 & 0.139 & 0.08/0.01  & 19.01/0.14  & -11.21/0.03  & 0.52/0.08  & 25.77/0.08  & -11.87/0.06  & 45.67  & 2.03  & 1.92  \\
J0238-312 & 1RXSJ023832.6-3116 & 0.232 & 0.15/0.01  & 16.46/0.16  & -11.10/0.04  &  &  &  &  & 1.84  & 2.67  \\
J0303-241 & PKS0301-243 & 0.26 & 0.20/0.01  & 15.22/0.09  & -10.90/0.03  & 0.63/0.18  & 24.02/0.18  & -10.74/0.15  & 46.67  & 1.92  & 2.49  \\
J0316+413 & IC310 & 0.019 & 0.25/0.03  & 14.59/0.17  & -10.69/0.19  &  &  &  &  & 1.90  &  \\
J0319+187 & RBS0413 & 0.19 & 0.12/0.01  & 17.35/0.19  & -11.33/0.03  & 0.39/0.25  & 24.61/0.28  & -11.55/0.17  & 45.88  & 1.57  & 2.20  \\
J0349-119 & 1ES0347-121 & 0.188 & 0.10/0.01  & 18.77/0.44  & -11.07/0.07  & 0.59/0.15  & 25.31/0.11  & -11.79/0.11  & 46.08  & 1.73  & 2.01  \\
J0416+010 & 1ES0414+009 & 0.287 & 0.13/0.01  & 16.77/0.16  & -11.15/0.03  & 0.73/0.20  & 24.50/0.15  & -11.58/0.16  & 46.42  & 1.75  & 2.38  \\
J0449-438 & PKS0447-439 & 0.205 & 0.22/0.01  & 15.46/0.13  & -10.52/0.05  & 0.62/0.14  & 24.16/0.13  & -10.50/0.13  & 46.73  & 1.85  & 2.70  \\
J0507+676 & 1ES0502+675 & 0.34 & 0.13/0.01  & 17.64/0.16  & -10.77/0.03  & 0.39/0.11  & 25.19/0.23  & -10.96/0.10  & 47.06  & 1.52  & 2.30  \\
J0509+056 & TXS0506+056 &  & 0.25/0.01  & 14.40/0.10  & -10.79/0.04  & 0.09/0.02  & 22.89/0.48  & -11.00/0.09  &  & 2.04  & 2.32  \\
J0521+211 & VERJ0521+211 & 0.108 & 0.18/0.01  & 15.50/0.11  & -10.69/0.03  & 0.24/0.11  & 23.50/0.52  & -10.65/0.11  & 46.03  & 1.92  & 2.35  \\
J0550-322 & PKS0548-322 & 0.069 & 0.10/0.02  & 17.92/0.60  & -10.93/0.07  & 0.61/0.08  & 25.23/0.06  & -11.85/0.06  & 45.23  & 1.61  & 1.86  \\
J0648+152 & RXJ0648.7+1516 & 0.179 & 0.15/0.01  & 16.80/0.21  & -11.03/0.06  & 0.46/0.23  & 24.76/0.17  & -11.26/0.15  & 46.08  & 1.83  & 2.36  \\
J0650+250 & 1ES0647+250 & 0.203 & 0.16/0.01  & 16.33/0.18  & -10.89/0.05  &  &  &  &  & 1.72  & 2.33  \\
J0710+591 & RGBJ0710+591 & 0.125 & 0.10/0.01  & 18.34/0.55  & -10.94/0.07  & 0.26/0.08  & 24.84/0.16  & -11.63/0.10  & 45.81  & 1.66  & 1.79  \\
J0721+713 & S50716+714 & 0.127 & 0.27/0.01  & 14.43/0.04  & -10.19/0.04  & 0.12/0.02  & 22.72/0.21  & -10.33/0.11  & 46.63  & 2.04  & 2.18  \\
J0739+016 & PKS0736+01$^*$ & 0.189 & 0.26/0.01  & 13.66/0.09  & -10.98/0.03  & 0.23/0.02  & 20.98/0.06  & -10.47/0.12  & 46.58  & 2.48  & 1.70  \\
J0809+523 & 1ES0806+524 & 0.138 & 0.17/0.01  & 15.91/0.12  & -10.93/0.03  & 0.22/0.12  & 24.31/0.28  & -11.09/0.11  & 45.96  & 1.88  & 2.49  \\
J0847+115 & RBS0723 & 0.199 & 0.10/0.01  & 18.54/0.23  & -11.22/0.07  & 0.29/0.28  & 25.27/1.07  & -11.70/0.12  & 46.03  & 1.74  & 1.93  \\
J0854+201 & OJ287 & 0.306 & 0.28/0.01  & 13.76/0.07  & -10.41/0.05  & 0.12/0.02  & 21.94/0.22  & -10.86/0.09  & 47.19  & 2.18  & 1.72  \\
J0958+655 & S40954+65 & 0.367 & 0.34/0.02  & 13.38/0.08  & -10.98/0.05  & 0.09/0.01  & 21.37/0.12  & -11.29/0.08  & 46.88  & 2.38  & 1.69  \\
J1010-313 & 1RXSJ101015.9-3119 & 0.143 & 0.12/0.02  & 17.47/0.49  & -11.01/0.10  & 0.29/0.15  & 24.12/0.42  & -11.54/0.17  & 45.86  & 1.58  & 2.33  \\
J1015+494 & 1ES1011+496 & 0.212 & 0.15/0.01  & 16.14/0.17  & -10.79/0.04  & 0.26/0.10  & 24.61/0.16  & -10.75/0.09  & 46.62  & 1.83  & 2.54  \\
J1103-234 & 1ES1101-232 & 0.186 & 0.12/0.01  & 17.53/0.26  & -10.82/0.05  & 0.28/0.13  & 24.92/0.21  & -11.68/0.14  & 46.28  & 1.64  & 2.05  \\
J1104+382 & mrk421 & 0.031 & 0.14/0.01  & 17.38/0.31  & -9.71/0.07  & 0.44/0.06  & 24.87/0.09  & -9.72/0.10  & 45.81  & 1.77  & 2.29  \\
J1136+676 & RXJ1136.5+6737 & 0.136 & 0.10/0.01  & 18.15/0.61  & -11.20/0.08  &  &  &  &  & 1.72  & 1.92  \\
J1136+701 & Mrk180 & 0.045 & 0.16/0.01  & 15.91/0.21  & -10.89/0.08  & 0.29/0.09  & 24.82/0.15  & -11.39/0.09  & 44.84  & 1.82  & 2.38  \\
J1159+292 & TON0599$^*$ & 0.725 & 0.23/0.02  & 13.69/0.11  & -11.15/0.07  & 0.17/0.02  & 22.08/0.15  & -10.61/0.07  & 48.02  & 2.21  & 1.68  \\
J1217+301 & 1ES1215+303 & 0.13 & 0.22/0.01  & 14.81/0.05  & -10.88/0.04  & 0.32/0.10  & 23.87/0.30  & -10.89/0.10  & 45.95  & 1.97  & 2.65  \\
J1221+282 & WComae & 0.103 & 0.23/0.01  & 14.66/0.07  & -10.69/0.04  & 0.12/0.03  & 21.96/0.22  & -10.62/0.16  & 46.06  & 2.10  & 2.41  \\
J1221+301 & 1ES1218+304 & 0.182 & 0.15/0.01  & 16.87/0.15  & -10.86/0.04  & 0.62/0.09  & 24.75/0.07  & -10.85/0.09  & 46.32  & 1.66  & 2.14  \\
J1224+213 & 4C+21.35$^*$ & 0.435 & 0.21/0.01  & 14.30/0.11  & -10.94/0.05  & 0.22/0.04  & 22.29/0.21  & -9.98/0.14  & 47.93  & 2.29  & 1.57  \\
J1224+246 & MS1221.8+2452 & 0.219 & 0.19/0.02  & 15.76/0.34  & -11.47/0.08  &  &  &  &  & 1.89 &  \\
J1230+253 & S31227+25 & 0.135 & 0.25/0.02  & 14.52/0.14  & -10.89/0.05  & 0.48/0.27  & 23.41/0.38  & -11.56/0.11  & 45.76  & 2.24  &  \\
J1256-057 & 3C279$^*$ & 0.536 & 0.27/0.02  & 13.14/0.10  & -10.44/0.06  & 0.12/0.02  & 22.31/0.17  & -10.16/0.09  & 48.21  & 2.34  & 1.60  \\
J1315-426 & 1ES1312-423 & 0.105 & 0.15/0.03  & 17.04/0.57  & -11.17/0.14  &  &  &  &  & 2.08  & 2.11  \\
J1427+238 & PKS1424+240 & 0.16 & 0.22/0.01  & 15.30/0.10  & -10.53/0.04  & 0.98/0.17  & 24.15/0.09  & -10.30/0.14  & 46.51  & 1.82  & 2.64  \\
J1428+426 & H1426+428 & 0.129 & 0.11/0.01  & 17.99/0.40  & -10.89/0.06  & 0.56/0.40  & 24.87/0.35  & -11.48/0.22  & 45.86  & 1.57  & 2.06  \\
J1443+120 & 1ES1440+122 & 0.163 & 0.11/0.01  & 17.50/0.26  & -11.33/0.05  & 0.41/0.16  & 24.85/0.20  & -11.76/0.09  & 45.68  & 1.80  & 2.06  \\
J1443+250 & PKS1441+25$^*$ & 0.939 & 0.18/0.04  & 13.49/0.32  & -12.50/0.10  & 0.13/0.02  & 22.13/0.21  & -11.58/0.08  & 47.38  & 2.13  &  \\
J1443-391 & PKS1440-389 & 0.065 & 0.19/0.01  & 15.71/0.14  & -10.86/0.04  & 0.22/0.06  & 24.73/0.17  & -11.14/0.04  & 45.25  & 1.81  & 2.64  \\
J1512-091 & PKS1510-089$^*$ & 0.36 & 0.26/0.02  & 13.56/0.11  & -10.92/0.04  & 0.21/0.02  & 21.77/0.08  & -9.79/0.09  & 47.90  & 2.36  & 1.28  \\
J1517-243 & APLibrae & 0.048 & 0.28/0.01  & 13.81/0.09  & -10.62/0.03  & 0.10/0.01  & 21.67/0.07  & -10.84/0.04  & 45.25  & 2.11  & 1.61  \\
J1555+111 & PG1553+113 & 0.36 & 0.19/0.01  & 15.70/0.06  & -10.43/0.03  & 0.84/0.11  & 24.41/0.06  & -10.21/0.11  & 47.50  & 1.68  & 2.42  \\
J1653+397 & mrk501 & 0.034 & 0.11/0.01  & 16.83/0.20  & -10.35/0.05  & 0.31/0.06  & 24.82/0.09  & -10.57/0.08  & 45.23  & 1.72  & 2.11  \\
J1725+118 & H1722+119 & 0.018 & 0.14/0.02  & 16.70/0.44  & -10.95/0.08  & 0.15/0.14  & 23.75/0.74  & -11.16/0.10  & 44.08  & 1.89  & 2.74  \\
J1728+502 & 1ES1727+502 & 0.055 & 0.13/0.01  & 16.53/0.28  & -11.12/0.05  &  &  &  &  & 1.96  & 2.28  \\
J1743+196 & 1ES1741+196 & 0.084 & & & & & & &  & 1.78  & 2.17  \\
J1751+096 & OT081 & 0.322 & 0.30/0.04  & 13.52/0.19  & -10.75/0.07  & 0.12/0.02  & 21.31/0.11  & -10.74/0.08  & 47.13  & 2.25  & 1.79  \\
J1943+213 & HESSJ1943+213 &  & 0.08/0.01  & 19.25/0.52  & -11.03/0.06  &  &  &  &  &  &  \\
J1959+651 & 1ES1959+650 & 0.047 & 0.15/0.01  & 16.82/0.17  & -10.32/0.05  & 0.22/0.05  & 24.73/0.15  & -10.78/0.08  & 45.48  & 1.88  & 2.32  \\
J2001+438 & MAGICJ2001+435 &  & 0.20/0.01  & 15.34/0.13  & -11.00/0.06  & 0.33/0.20  & 23.76/0.30  & -10.73/0.10  &  & 1.97  & 2.54  \\
J2009-488 & PKS2005-489 & 0.071 & 0.14/0.01  & 16.33/0.17  & -10.12/0.06  & 0.45/0.05  & 24.36/0.09  & -10.99/0.07  & 45.99  & 1.77  & 2.62  \\
J2039+523 & 1ES2037+521 & 0.053 & 0.16/0.02  & 15.92/0.37  & -11.56/0.13  &  &  & &  & 1.89  & 2.62  \\
J2056+496 & RGBJ2056+496 &  & 0.18/0.02  & 15.94/0.22  & -10.57/0.13  & & & &  & 1.78  & 2.42  \\
J2158-302 & PKS2155-304 & 0.116 & 0.21/0.01  & 15.73/0.07  & -9.93/0.04  & 0.39/0.10  & 24.69/0.15  & -9.90/0.16  & 46.79  & 1.83  & 2.64  \\
J2202+422 & BLLac & 0.069 & 0.27/0.02  & 13.90/0.10  & -10.24/0.03  & 0.09/0.01  & 21.84/0.23  & -10.49/0.07  & 45.97  & 2.25  & 1.79  \\
J2243+203 & RGBJ2243+203 &  & 0.20/0.01  & 15.54/0.10  & -11.03/0.04  & 0.24/0.23  & 24.21/0.39  & -11.13/0.14  &  & 1.79  & 2.76  \\
J2250+384 & B32247+381 & 0.119 & 0.11/0.01  & 17.47/0.56  & -11.18/0.09  & 0.21/0.15  & 24.20/0.39  & -11.53/0.14  & 45.56  & 1.91  & 2.43  \\
J2347+517 & 1ES2344+514 & 0.044 & 0.12/0.01  & 16.90/0.24  & -11.00/0.05  & 0.31/0.13  & 25.22/0.22  & -10.93/0.19  & 44.91  & 1.78  & 2.09  \\
J2359-306 & H2356-309 & 0.165 & 0.11/0.01  & 18.02/0.38  & -10.99/0.04  & 0.19/0.11  & 24.19/0.41  & -11.72/0.12  & 46.03  & 2.02  & 2.07  \\
\end{longtable}
\end{small}
\end{center}
\end{landscape}

\begin{table}
\begin{center}
\caption[]{Statistical Values for the Fitting Results.}
 \begin{tabular}{ccccc}
  \hline\noalign{\smallskip}
Parameter &
min &
max &
mean &
std                    \\
  \hline\noalign{\smallskip}
    $|2c^{s}|$   & 0.08 & 0.41 & 0.18 & 0.07\\
    $\log \nu^{s}_{p}$  & 12.85 & 19.31 & 16.07 & 1.62\\
    $\log \nu^{s}_{\rm p}f_{\nu^{s}_{\rm p}}$  & $-$12.21 & $-$9.70 & $-$10.81 & 0.39\\
    $|2c^{IC}|$   & 0.08 & 0.98 & 0.33 & 0.20\\
    $\log \nu^{IC}_{p}$  & 21.05 & 25.83 & 23.90 & 1.24\\
    $\log \nu^{IC}_{\rm p}f_{\nu^{IC}_{\rm p}}$  & $-$11.82 & $-$9.66 & $-$10.89 & 0.57\\
    $\log L_{bol}$ & 44.08 & 48.21 & 46.31 & 0.93\\
  \noalign{\smallskip}\hline
\end{tabular}
\end{center}
\end{table}

\subsection{Curvature and Peak Frequency}

\begin{figure}[h]
\centering
\includegraphics[width=3.0 in]{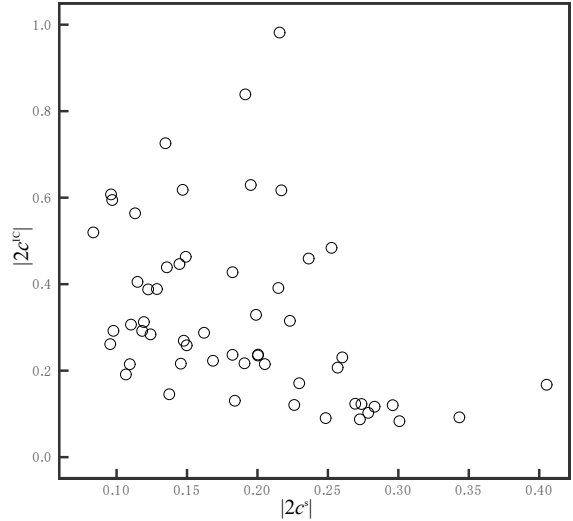}
\caption{\label{ms0075fig3} Plot of curvature of inverse-Compton bump against that of synchrotron bump. }
\end{figure}

From our fitting results,
    we plot the curvature of IC bump against that of the synchrotron bump in Figure \ref{ms0075fig3}.
Based on a T-test,
    we find that the curvature of IC bump is,
    on average,
    higher than that of synchrotron bump with a chance probability $p<10^{-4}$.
A Spearman's rank correlation indicates that there is a significant anti-correlation
    between synchrotron curvature and IC curvature with $p = 4.28\times10^{-4}$.
However, there are only 9 sources with $|2c^{s}| > 0.26$, which dominate the correlation analysis,
    and the scatter of the correlation is large.
In this sense, we need more sources with $|2c^s| > 0.26$ to investigate such an anti-correlation.

\begin{figure}[h]
\centering
\includegraphics[width=3.0 in]{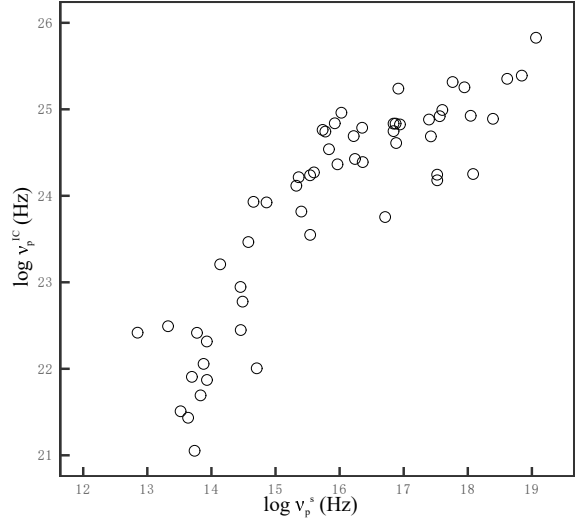}
\caption{\label{ms0075fig4} Plot of peak frequency of inverse-Compton bump against that of synchrotron bump.}
\end{figure}

A strong correlation is found between the two bump peak frequencies,
  namely $\log \nu^{IC}_{p} = (0.66 \pm 0.05)\log \nu^{s}_{p} + (13.38 \pm 0.81)$
  with a correlation coefficient $r = 0.87$ and a chance probability $p < 10^{-4}$,
  see Fig \ref{ms0075fig4}.
Because the scatter of the $\log \nu^{IC}_{p}-\log \nu^{s}_{p}$ correlation is about one dex,
and $\log \nu^{IC}_{p}$ is hard to be obtained for some TeV sources,
    we suggest to use $\log \nu^{s}_{p}$ for its estimation.

\begin{figure}[h]
\centering
\includegraphics[width=3.0 in]{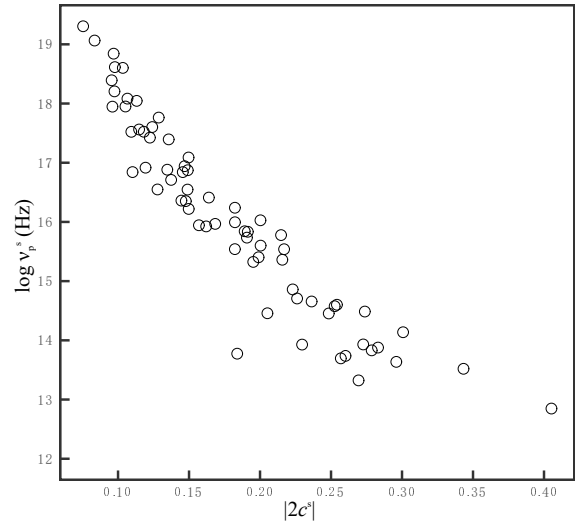}
\caption{\label{ms0075fig5} Plot of peak frequency against curvature for synchrotron bump.}
\end{figure}

\begin{figure}[h]
\centering
\includegraphics[width=3.0 in]{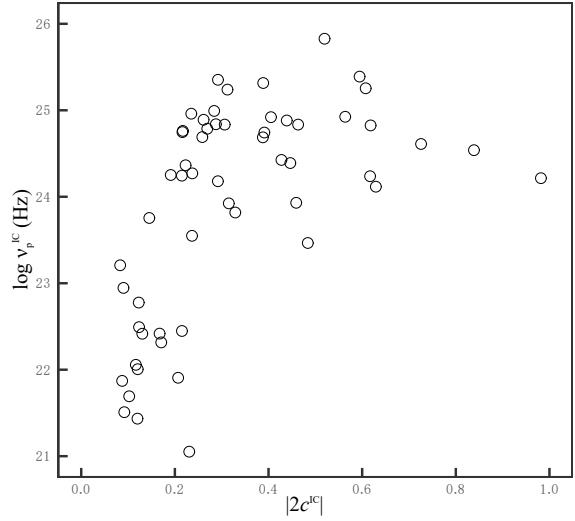}
\caption{\label{ms0075fig6} Plot of peak frequency against curvature for inverse-Compton bump.}
\end{figure}

For the correlation between $\log \nu_{p}$ and $|2c|$,
    we have  $\log \nu^{s}_{p} = -(21.87 \pm 1.04)|2c^{s}| + (20.01 \pm 0.20)$ with
    $r = -0.93$ and $p < 10^{-4}$,
    a Spearman's rank correlation indicates that there is a strongly positive correlation
     between $\log \nu^{IC}_{p}$ and $|2c^{IC}|$ with $r = 0.63$ and $p<10^{-4}$,
     see Figures \ref{ms0075fig5} and \ref{ms0075fig6}.

\subsection{X/$\gamma$-ray Photon Index and Peak Frequency}

\begin{figure}[h]
\centering
\includegraphics[width=3.0 in]{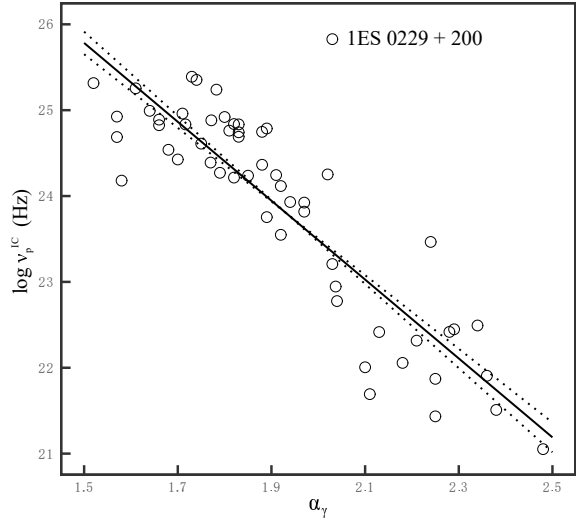}
\caption{\label{ms0075fig7} Plot of inverse-Compton peak frequency against $\gamma$-ray photon index. The full line stands for the fitting result ($y=ax+b$) which exclude the source 1ES 0229+200. The dash lines stand for the fitting result with 1 $\sigma$ error, namely $y=(a+\sigma)x+(b-\sigma$) and $y=(a-\sigma)x+(b+\sigma$).}
\end{figure}

When the inverse Compton peak frequency ($\log \nu^{IC}_{p}$ ) is plotted against the $\gamma$-ray spectral index ($\alpha_{\gamma}$),
there is a tendency for $\log \nu^{IC}_{p}$ to decrease with $\alpha_{\gamma}$ as shown in Figure \ref{ms0075fig7}.
 Excluding the source 1ES 0229+200,
    we find a close anti-correlation,
      $\log \nu^{IC}_{p} = -(4.59 \pm 0.30) \alpha_{\gamma} + (32.67 \pm 0.59)$ with $ r = -0.90$ and $p < 10^{-4}$.

\begin{figure}[h]
\centering
\includegraphics[width=3.0 in]{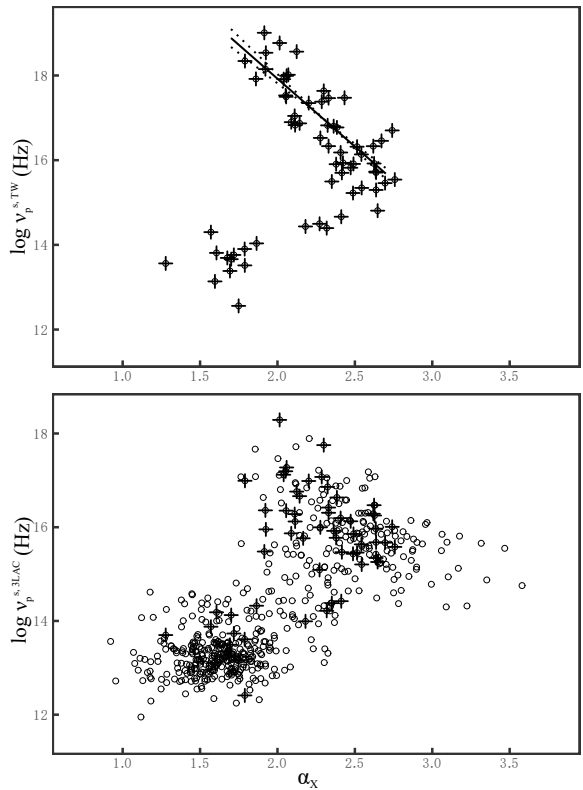}
\caption{\label{ms0075fig8}
  Plot of synchrotron peak frequency in this work against X-ray photon index for TeV blazars (upper panel),
  and synchrotron peak frequency in 3LAC
  (Ackermann et al. 2015) against X-ray photon index for Fermi blazars (lower panel).
  In the figure, circles stand for the Fermi blazars, and filled circles with cross stand for TeV Fermi blazars.
  The straight line stands for the fitting result ($y=ax+b$) for HSP TeV blazars. The dash lines stand for the fitting result with 1 $\sigma$ error, namely $y=(a+\sigma)x+(b-\sigma$) and $y=(a-\sigma)x+(b+\sigma$).}
\end{figure}

A clear "L-shape" relation is found between $\log \nu^{s}_{p}$ and $\alpha_X$ for TeV blazars,
    which implies that $\log \nu^{s}_{p}$ is highly correlated with $\alpha_X$, see the upper panel of Figure \ref{ms0075fig8}.
For the HSP TeV sources ($\log \nu_{p}^{s} > 15.3$ Hz), we have that $\log \nu^{s}_{p} = -(3.20 \pm 0.34) \alpha_X + (24.33 \pm 0.79)$
    with $r = -0.82$ and $p<10^{-4}$, and a positive correlation for LSP/ISP TeV sources.
Similar result is found for the whole Fermi blazars, see the lower panel of Figure \ref{ms0075fig8}.
There is a positive correlation between $\log \nu^{s}_{p}$ and $\alpha_X$ for ISP/LSP blazars,
    while there is an anti-correlation between them for HSP blazars.
From Figure \ref{ms0075fig8}, we found that the changing point of the "L-shape" relation is around $\log \nu_{p}^{s}=15$ Hz, which is very close to boundary between HSP and ISP determined by Abdo et al.(2010) and Fan et al.(2016). Does that mean that we can classify TeV blazars as HSP-TeV and LSP-TeV blazars using log $\nu^s_{\rm peak} (\rm Hz) = 15$?

\begin{figure}[h]
\centering
\includegraphics[width=3.0 in]{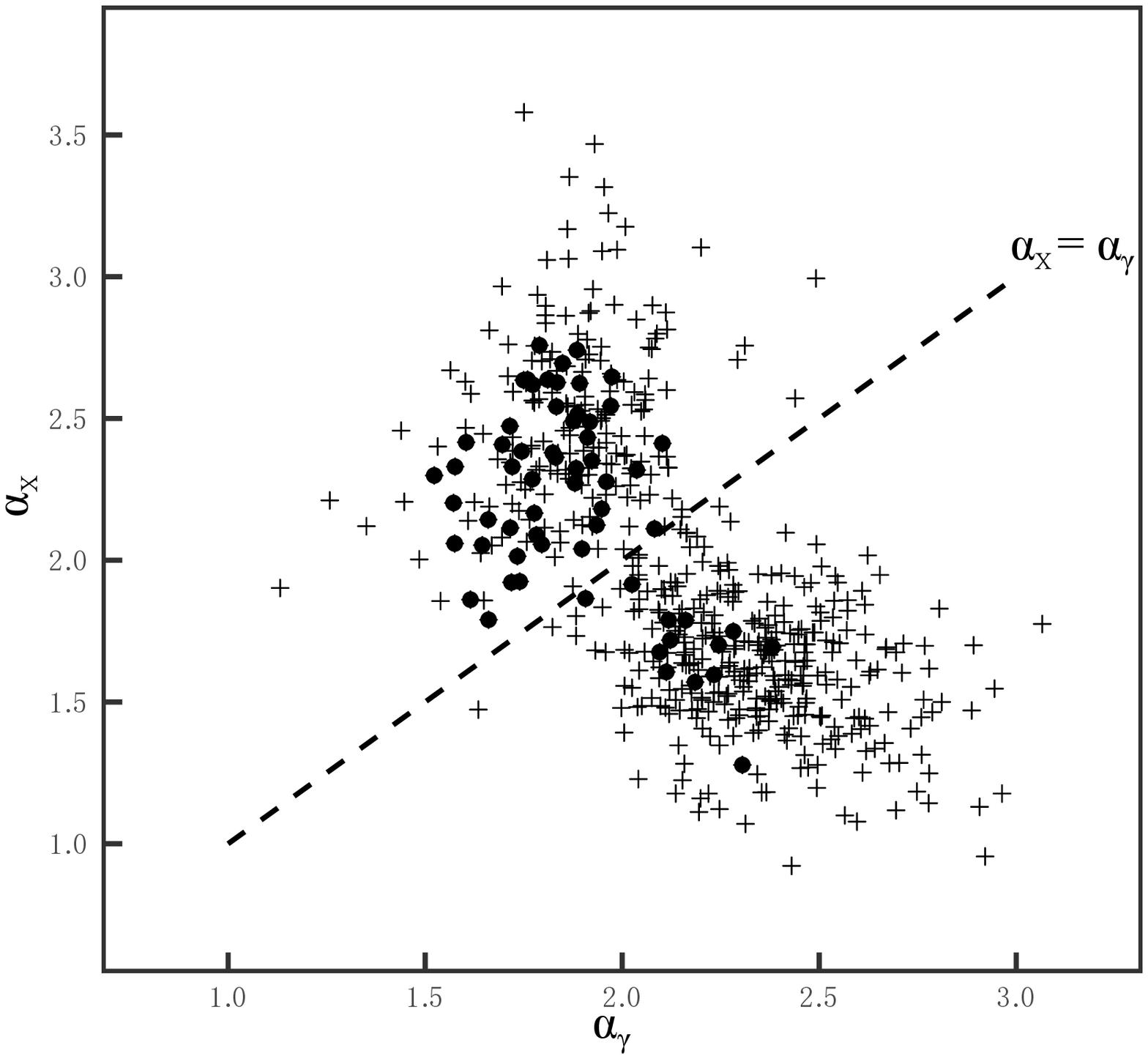}
\caption{\label{ms0075fig9}
  Plot of X-ray photon index against $\gamma$-ray photon index for TeV blazars (filled point symbols),
  and that for Fermi blazars (plus symbols). The straight dash line stands for $\alpha_X = \alpha_\gamma$.}
\end{figure}

Another "L-shape" relation is found between $\alpha_X$ and $\alpha_{\gamma}$ for TeV blazars.
 As $\alpha_{\gamma}$ increases $\alpha_X$ increases and then decrease,
  see dot symbols of Figure \ref{ms0075fig9}.
When we plot the whole Fermi blazars sample in $\alpha_X-\alpha_{\gamma}$ panel,
  two separated sub-samples are also obvious,
  and we can simply use a straight line ($\alpha_X=\alpha_{\gamma}$) to separate the two sub-samples.
TeV blazars mainly occupy the upper-left part of the whole Fermi blazars sample in Figure \ref{ms0075fig9},
    namely they have softer X-ray and harder $\gamma$-ray spectral index than others.
The lower-right part sample shows a tendency anti-correlation between $\alpha_X$ and $\alpha_{\gamma}$ with $r =-0.28$ and $p<10^{-4}$,
 while the upper-left part sample shows a tendency positive correlation between them with $r = 0.29$ and $p<10^{-4}$, see Figure \ref{ms0075fig9}.
Our result of the anti-correlation between $\alpha_X$ and $\alpha_{\gamma}$ is consistent with those in the work of
 Wang et al. (1996) and our previous work (Fan et al. 2012).
The positive correlation in the present work was not obtained in our previous work because of the sample in our previous work is limited.

\subsection{Luminosity-Luminosity Correlation}

\begin{figure}[h]
\centering
\includegraphics[width=3.0 in]{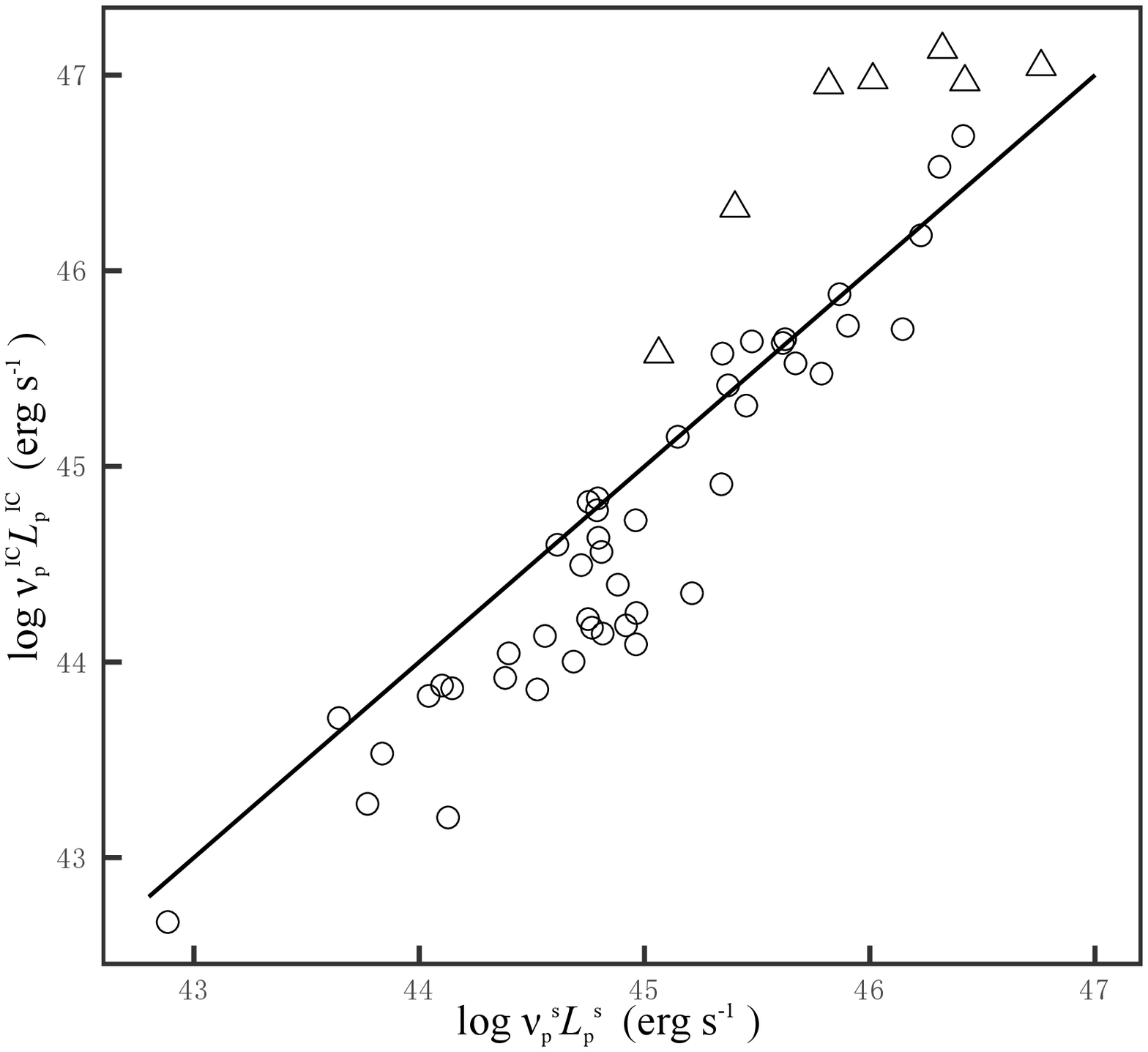}
\caption{\label{ms0075fig10} Plot of IC peak luminosity against synchrotron peak luminosity. The straight line stands for $\log \nu_p^s L_p^s = \log \nu_p^{IC} L_p^{IC}$. The circle symbol stands for TeV BL Lacs, and the triangle symbol stands for TeV FSRQs.}
\end{figure}

For luminosity and luminosity correlation, we find that there are strong correlations between peak luminosities and  bolometric luminosity, they are:

$\log\nu_{p}^{IC} L _{p}^{IC} = (1.28 \pm 0.07) \log\nu_{p}^{s} L _{p}^{s} - (12.64 \pm 3.21)$ with $r=0.93$ and $p<10^{-4}$ for 52 TeV blazars, see Figure \ref{ms0075fig10};

\begin{figure}[h]
\centering
\includegraphics[width=3.0 in]{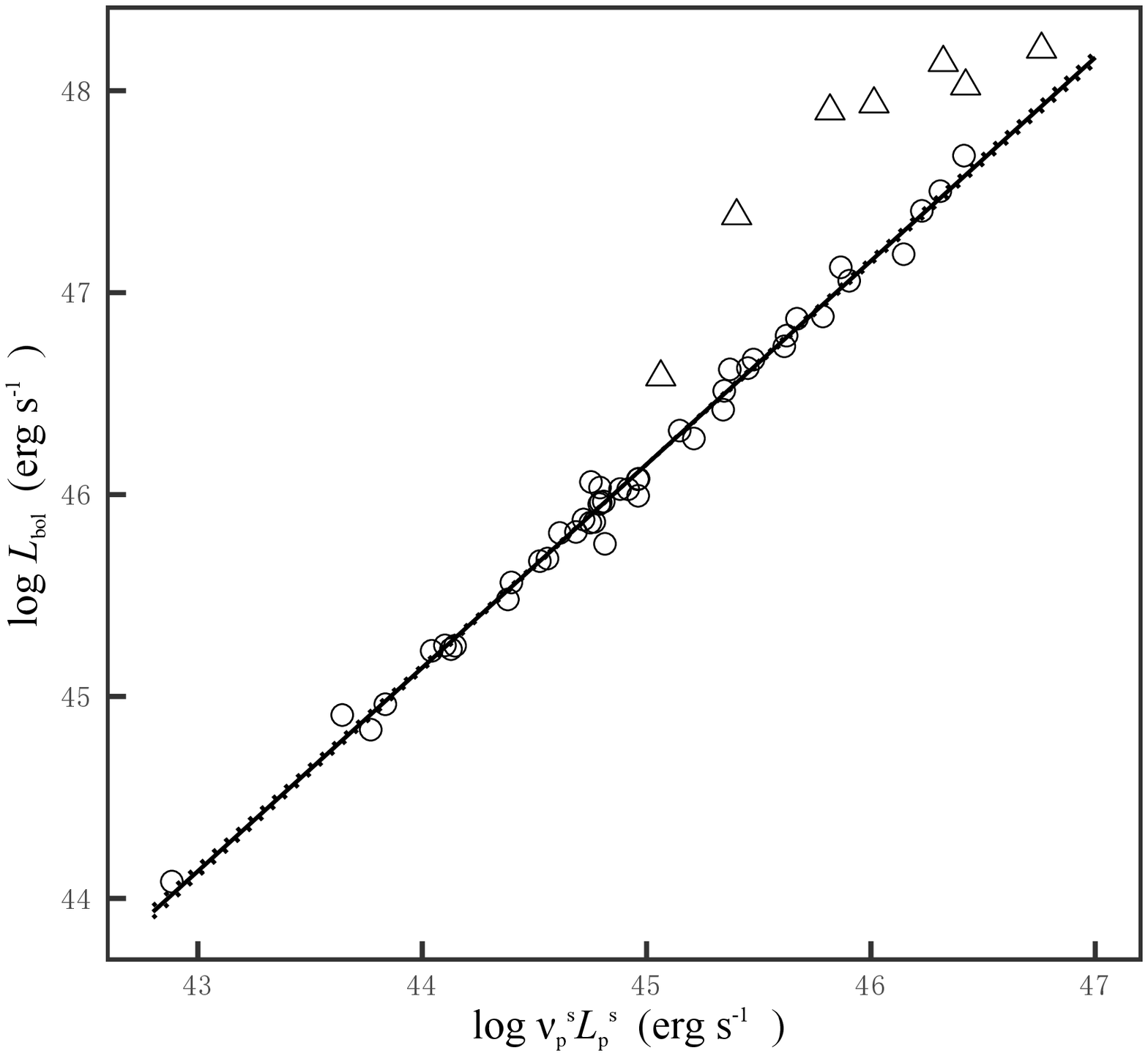}
\caption{\label{ms0075fig11} Plot of bolometric luminosity against synchrotron peak luminosity. The circle symbol stands for TeV BL Lacs, and the triangle symbol stands for TeV FSRQs. The straight line stands for the fitting result ($y=ax+b$) for TeV BL Lacs. The dash lines stand for the fitting result with 1 $\sigma$ error, namely $y=(a+\sigma)x+(b-\sigma$) and $y=(a-\sigma)x+(b+\sigma$).}
\end{figure}

$\log L _{bol} = (1.01 \pm 0.01) \log\nu_{p}^{s} L _{p}^{s} + (0.82 \pm 0.62)$ with $r=0.996$ and $p<10^{-4}$ for 45 TeV BL Lacs, see Figure \ref{ms0075fig11};

\begin{figure}[h]
\centering
\includegraphics[width=3.0 in]{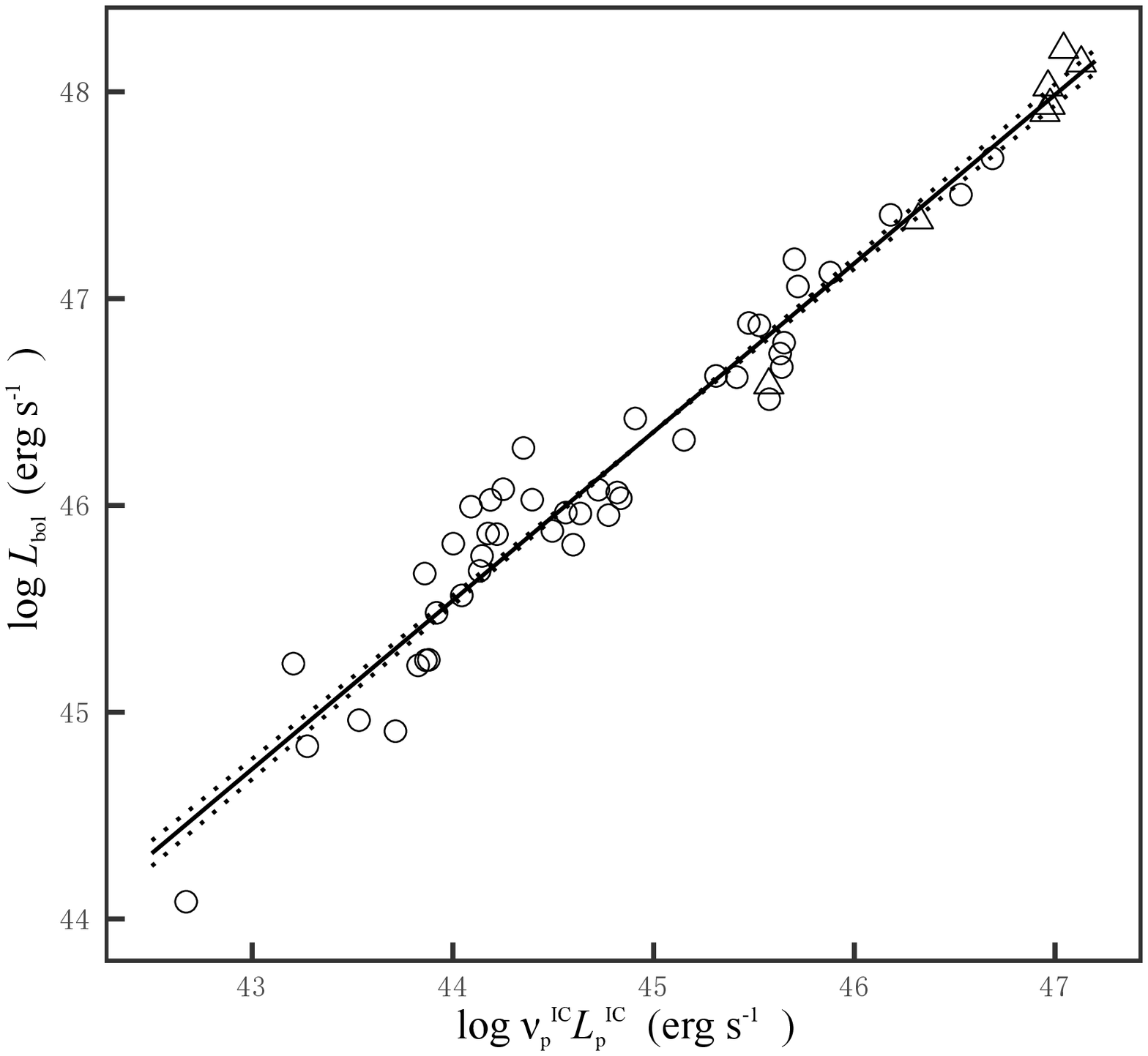}
\caption{\label{ms0075fig12} Plot of bolometric luminosity against IC peak luminosity. The circle symbol stands for TeV BL Lacs, and the triangle symbol stands for TeV FSRQs. The straight line stands for the fitting result ($y=ax+b$). The dash lines stand for the fitting result with 1 $\sigma$ error, namely $y=(a+\sigma)x+(b-\sigma$) and $y=(a-\sigma)x+(b+\sigma$).}
\end{figure}

$\log L _{bol} = (0.82 \pm 0.03) \log\nu_{p}^{IC} L _{p}^{IC} + (9.67 \pm 1.13)$ with $r=0.98$ and $p<10^{-4}$ for 52 TeV blazars, see Figure \ref{ms0075fig12}.

Figure \ref{ms0075fig10} shows that $\log\nu_{p}^{IC} L _{p}^{IC}$ and $\log\nu_{p}^{s} L _{p}^{s}$ are closely correlated with each other,
and $\log\nu_{p}^{s} L _{p}^{s}$ is higher than $\log\nu_{p}^{IC} L _{p}^{IC}$ for most of TeV BL Lacs, while  TeV FSRQs seems to have stronger IC emissions than synchrotron emissions.
From Figure \ref{ms0075fig11}, we find that FSRQs are the outliers of the $\log L _{bol}-\log\nu_{p}^{s} L _{p}^{s}$ correlation. So we exclude them when we analyzed the $\log L _{bol}-\log\nu_{p}^{s} L _{p}^{s}$ correlation.

\begin{figure}[h]
\centering
\includegraphics[width=3.0 in]{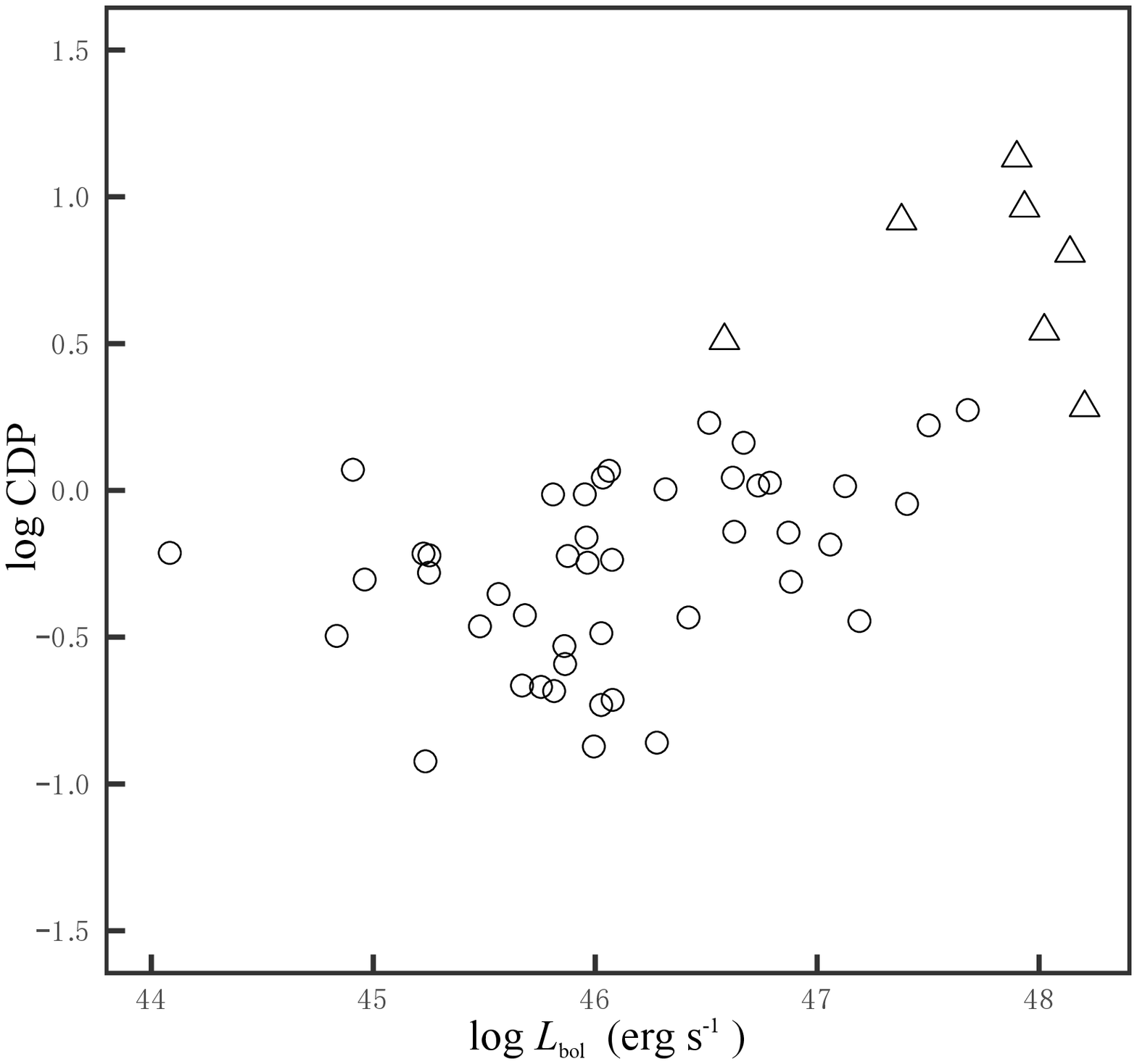}
\caption{\label{ms0075fig13} Plot of Compton dominance parameter against bolometric luminosity. The circle symbol stands for TeV BL Lacs, and the triangle symbol stands for TeV FSRQs.}
\end{figure}

In this work, the Compton dominance parameter ($CDP = \nu_{p}^{IC} L _{p}^{IC}/\nu_{p}^{s} L _{p}^{s}$) is calculated.
Then, the correlation between the bolometric luminosity and CDP is investigated,
and a strong correlation, $\log L _{bol} = (1.28 \pm 0.22) CDP + (46.48 \pm 0.10)$ with
  $r = 0.64$ and $p<10^{-4}$ is found,
  see Figure \ref{ms0075fig13}.
The sources with higher CDPs tend to have higher bolometric luminosity, and FSRQs locate in high $\log L _{bol}$  and high CDPs area.
Therefore, the synchrotron emissions are dominant in the bolometric emissions for the most weak power sources,
  while the IC emissions contribute the main part of power for bright sources.

\section{Discussions}

\subsection{Candidates for TeV Blazars}

Some researchers proposed TeV blazar candidates based on some selection criteria,
 "high peak frequency" is a very common one
(eg,. Massaro et al. 2013;
 Chang et al. 2017).
However, from Table 2, we found that TeV blazars cover a wide range of synchrotron peak frequency,
from $\rm log \nu_p$ = 12.85 Hz to
     $\rm log \nu_p$ = 19.31 Hz.
In our sample, 30.4\% of blazars are classified as ISP/LSP blazars following our classification criteria
  (Fan et al. 2016).
Thus, if "high peak frequency" are selected as one of the selection criteria of TeV blazar candidates,
more than a quarter of TeV sources will be ignored.

From Figure \ref{ms0075fig9}, we find that TeV blazars mainly occupy in the left-upper part of the whole Fermi blazars sample.
From our previous work (Lin \& Fan 2016) and the present work, we found that it is hard to completely separate TeV and non-TeV blazars.
In addition,
  TeV blazars have lower averaged redshift,
  higher averaged flux, higher peak frequency,
  flat $\gamma$-ray spectrum,
  stronger $\gamma$-ray emissions than non-TeV ones.
It is possible that all blazars are the emitters of TeV emissions,
 and TeV blazars are the sources with high TeV flux at certain times.

\subsection{ Curvature and Peak Frequency}

The basic structure of SEDs for blazars can reveal both the radiative
and acceleration mechanisms.
Tramacere et al. (2007, 2009) analyzed the SEDs of Mrk 421, they found the logarithm of peak frequency to be inversely proportional to the curvature, which indicate the electron energies under stochastic and systematic acceleration.
Our recent work (Fan et al. 2016) found a clear anti-correlation between synchrotron peak frequency and curvature for a sample of Fermi blazars
(see also Massaro et al. 2004b).
In this work, we also find a strong anti-correlation between $\log \nu_{p}^{s}$ and $|2c^{s}|$ for TeV balzars.
Thus, our results confirm those results and support the suggestion that the synchrotron SED results from a stochastic and systematic component in the acceleration process.
But, the $\log \nu_{p}-|2c|$ correlations are clearly different for the two bumps, which further indicates that there are different emission mechanisms between them.

A strong positive correlation is found between the two bump peak frequencies, but the Figure \ref{ms0075fig4} shows that this correlation seems to be a line with a small curved. That means $\log \nu_{p}^{s}-\log \nu_{p}^{IC}$ correlation is a little different between high peak sources and low peak sources. The strong correlation between $\log \nu_{p}^{s}$ and $\log \nu_{p}^{IC}$ supports the one-zone leptonic model, indicating that the two bumps are produced by the same electronic population.

\subsection{X/$\gamma$-ray Photon Index and Peak Frequency}

Abdo et al. (2010b) found that the inverse Compton peak frequency is correlated with the $\gamma$-ray spectral index as
  $\log \nu^{IC}_{p} = -4.0 \alpha_{\gamma} + 31.6$ for 48 Fermi bright blazars.
  Since the SEDs fitting results of their work were based on some quasi-simultaneous SEDs, they suggested that the above equation can be used to estimate the $\log \nu^{IC}_{p}$ of the sources which have no simultaneous SEDs.
In this work, we found $\log \nu^{IC}_{p} = -4.59 \alpha_{\gamma} + 32.67$, which is consistent with the result in Abdo et al. (2010b).
Since the detected wavelength of Fermi (1-100GeV) is in an important part of IC bump for blazars,
thus the $\alpha_{\gamma}$ is an indicator of $\log \nu^{IC}_{p}$ for blazars.

An "L-shape" relation is found between $\log \nu^{s}_{p}$ and $\alpha_X$ in this work for both TeV blazars and Fermi blazars.
Abdo et al. (2010b) discussed the behavior between $\log \nu^{s}_{p}$ and $\alpha_X$ for 48 fermi bright blazars,
 and suggested that the X-ray spectrum will be flat in the LSP sources and the extreme HSP sources.
However, there are only a few sources with $\log \nu _{p}^{s} > 10^{16}$Hz in their sample.
Thus, the result of our sample, which covers a wide range of $\log \nu^{s}_{p}$, confirms their results.
When the peak frequency moves to the lower frequency,
  the X-ray emissions are composed of the synchrotron tail and the inverse Compton emission resulting in
  a flat X-ray spectrum index, while the peak frequency is greater than $\rm 10^{15} Hz$ and move to the higher
  frequency, then the X-ray emission is mainly from the synchrotron emission and results in flat X-ray spectrum
  again.
Therefore, we have an "L-shape" relation between the X-ray spectrum index and the synchrotron peak frequency.
We also find that there is no clear difference between TeV and non-TeV sources in $\log \nu^{s}_{p}-\alpha_X$ diagram,
see lower panel of Figure \ref{ms0075fig8}.
A significant correlation between $\log \nu^{s}_{p}$ and $\alpha_X$ for HSP TeV blazars is also found.

Abdo et al. (2010b) found an  anti-correlation between $\alpha_\gamma$ and $\alpha_X$,
     and  proposed that such a correlation is excepted at first order in synchrotron-inverse Compton scenarios
(see also Wang et al 1996;
    Comastri et al. 1997;
    Fan et al. 2012;
    Ackermann et al. 2015).
Fan et al. (2012) found that the $\alpha_\gamma-\alpha_X$ correlation is not obvious in subclasses of blazars.
Figure \ref{ms0075fig9} shows an "L-shape" relation between $\alpha_X$ and $\alpha_\gamma$
 for both TeV blazars and Fermi blazars.
For the whole Fermi blazars sample, the $\alpha_X-\alpha_\gamma$ diagram looks like an anti-correlation.
However, when we investigated the TeV sub-samples separately,
 we find a
 positive correlation between $\alpha_X$ and $\alpha_\gamma$ at $\alpha_\gamma<2.0$ range with $r = 0.45$ and $p = 1.91 \times 10^{-3}$.
 The anti-correlation is consistent with those in the mentioned works,
    but the positive correlation was not reported before.

For the two separated sub-groups in Figure \ref{ms0075fig9},
    we find that the upper-left one corresponds to the sources with $\log \nu_p^s > 15.0$ Hz
  while the lower-right one corresponds to the sources with $\log \nu_p^s<15.0$ Hz.
Thus, the spectral indexes in X-ray and $\gamma$-ray bands are the indicators of the peak frequency of SEDs for blazars.
A similar result is also found in $\alpha_{RO}-\alpha_{OX}$ panel,
 in which HSP sources and LSP sources are located in the different parts of the panel
  (Abdo et al. 2010b, Fan et al. 2016, and the references therein).

\subsection{Bolometric Emissions}

In the $\log\nu_{p}^{s} L _{p}^{s}-\log\nu_{p}^{IC} L _{p}^{IC}$ correlation, we find that IC emissions are higher than synchrotron emissions
for strong sources whose $\log\nu_{p}^{s} L _{p}^{s} > 45 $ erg/s.
From Figures \ref{ms0075fig10} and \ref{ms0075fig11}, since $\log\nu_{p}^{IC} L _{p}^{IC}$ is larger than $\log\nu_{p}^{s} L _{p}^{s}$ for TeV FSRQs, the main part of bolometric luminosity should be contributed by IC emissions, which cause the FSRQs to be the outlier in $\log\nu_{p}^{s} L _{p}^{s}-\log\nu_{p}^{s} L _{p}^{s}$ correlatioin.
Some researchers also found that FSRQs has higher bolometric luminosity than BL Lasc (eg., M\"ucke et al. 2003; Giommi et al. 2012; Cha et al. 2014).
From Figure \ref{ms0075fig13}, it shows that the high power sources ($\log L _{bol} > 46.5 $ erg/s) have higher Compton dominance parameter ($\log CDP>0$).

From above investigation, we can see that the bolometric emissions are highly correlated with both the synchrotron emissions and the IC emissions, and the bolometric luminosity is mainly contributed by synchrotron emissions for BL Lacs, while that is contributed by EC emissions for FSRQs.
Therefore, we proposed that the bolometric luminosity can be estimated by the following equation,
$$
\log L_{bol} =\{
\begin{array}{l}
1.01\log\nu_{p}^{s} L _{p}^{s} + 0.82 \qquad \qquad {\rm for\, BL\, Lacs}\\
0.82\log\nu_{p}^{IC} L _{p}^{IC} + 9.67 \qquad {\rm for\, FSRQs}
\end{array}
$$
When we compared the
    estimated bolometric luminosity ($\log L _{bol}^{est}$) using above relations and
    the bolometric luminosity  from the SED fitting,
    we find that the difference between the two bolometric luminosity ($\log L _{bol}^{est} - \log L _{bol}$) is small, with mean value 0.12, standard deviation 0.07, and max value 0.33 for the sample in this work.

\section{Conclusions}
In this work, we collect a sample of 69 TeV blazars from TeVCat, obtain their observations from the SSDC SED Builder, and fit their SEDs by using the second degree polynomial function. The structure parameters of synchrotron bump for 68 blazars and those of IC bump for 56 blzars are obtained, the corresponding statistical values of some parameters, including curvature, peak frequency, peak luminosity, bolometric luminosity, and X/$\gamma$-ray spectral indexes are obtained.

Our main conclusions are as follows:

1. There is a clear positive correlation between
    the synchrotron peak frequency, $\log \nu_{\rm p}^{\rm s}$, and
    the inverse-Compton peak frequency $\log \nu_{\rm p}^{\rm IC}$, and that
    between the synchrotron peak luminosity, $\log \nu_{\rm p}^{\rm s} L_{p}^{\rm s}$, and
    the inverse-Compton peak luminosity, $\log \nu_{\rm p}^{\rm IC} L_{p}^{\rm IC}$.

2.  The correlation between the peak frequency and the curvature of synchrotron bump
    is clearly different from that of the inverse-Compton bump,
    which further indicates that there are different emission mechanisms between them.

3.  There is a correlation between $\log \nu_{\rm p}^{\rm IC}$ and $\gamma$-ray spectral index,
    $\alpha_{\gamma}$, for the TeV blazars:
    $\log \nu^{IC}_{p} = -(4.59 \pm 0.30) \alpha_{\gamma} + (32.67 \pm 0.59)$,
    which is consistent with previous work of Abdo et al.

4. An "L-shape" relation was found between $\log \nu^{s}_{p}$ and $\alpha_X$ for both TeV blazars and Fermi blazars.
    A significant correlation between $\log \nu^{s}_{p}$ and X-ray photon index ($\alpha_X$)
     is found for the TeV blazars with high synchrotron peak frequency (HSPs):
     $\log \nu^{s}_{p} = -(3.20 \pm 0.34) \alpha_X + (24.33 \pm 0.79)$, while the correlation is positive for LSP TeV sources.

5.  In the $\alpha_X-\alpha_\gamma$ diagram, there is  also "L-shape", the anti-correlation is consistent with the
available results in the literature, we also found a positive correlation between them.

6. Inverse-Compton dominant sources have luminous bolometric luminosities.

\section*{Acknowledgement}

The authors thank the referee for the useful comments and suggestions!
This work is supported by the National Natural Science Foundation of China (11733001, U1531245),
Guangdong Innovation Team (2014KCXTD014),
Nature Scientific Foundation of Guangdong Province (2017A030313011),
 and also supported from Astrophysics  Key Subjects of Guangdong Province and Guangzhou City.


\begin{thebibliography}{}
\bibitem[Aartsen et al. (2015)]{Aartsen15} Aartsen, M. G., Ackermann, M., Adams, J., et al., 2015, PhRvD, 91b2001
\bibitem[Abdo et al. (2010a)]{Abdo10a} Abdo, A. A., Ackermann, M., Ajello, M., et al., 2010a, ApJ, 715, 429
\bibitem[Abdo et al. (2010b)]{Abdo10b} Abdo, A. A., Ackermann, M., Agudo, I., et al., 2010b, ApJ, 716, 30
\bibitem[Abdo et al. (2014)]{Abdo14} Abdo, A. A., Abeysekara, A. U., Allen, B. T., et al., 2014, ApJ, 782, 110
\bibitem[Abramowski et al. (2013)]{Abramowski13} Abramowski, A., Acero, F., Aharonian, F., et al., 2013, PhRvD, 88j2003A
\bibitem[Abramowski et al. (2015)]{Abramowski15} Abramowski, A., Aharonian, F., Ait Benkhali, F., et al., 2015, ApJ, 802, 65
\bibitem[Acciari et al. (2008)]{Acciari08} Acciari, V. A., Aliu, E., Beilicke, M., et al., 2008, ApJ, 684L, 73
\bibitem[Acciari et al. (2011)]{Acciari11} Acciari, V. A., Aliu, E., Arlen, T., et al., 2011, ApJ, 738, 169
\bibitem[Acero et al. (2015)]{Acero15} Acero, F., Ackermann, M., Ajello, M., et al., 2015, ApJS, 218, 23
\bibitem[Ackermann et al. (2015)]{Acker15} Ackermann, M.; Ajello, M.; Atwood, W. B., et al., 2015, ApJ, 810, 14
\bibitem[Aharonian et al. (2001)]{Aharonian01} Aharonian, F. A., Akhperjanian, A. G., Barrio, J. A., et al., 2001, A\&A, 366, 62
\bibitem[Aharonian et al. (2003)]{Aharonian03} Aharonian, F., Akhperjanian, A., Beilicke, M., et al., 2003, A\&A, 406L, 9
\bibitem[Aharonian et al. (2009)]{Aharonian09} Aharonian, F., Akhperjanian, A. G., Anton, G., et al., 2009, ApJ, 696L, 150
\bibitem[Albert et al. (2008)]{Albert15} Albert, J., Aliu, E., Anderhub, H., et al.,  2008, Sci, 320, 1752
\bibitem[Aller et al. (1992)]{Aller92} Aller, M. F., Aller, H. D., Hughes, P. A., 1992, ApJ, 399, 16
\bibitem[Aliu et al. (2011)]{Aliu11} Aliu, E., Aune, T., Beilicke, M.,  et al., 2011, ApJ, 742, 127
\bibitem[Aliu et al. (2015)]{Aliu15} Aliu, E., Archer, A., Aune, T., et al., 2015, ApJ, 799, 7
\bibitem[Amenomori et al. (2003)]{Amenomori03} Amenomori, M., Ayabe, S., Cui, S. W.,  et al., 2003, ApJ, 598, 242
\bibitem[Archambault et al. (2013)]{Archambault13} Archambault, S., Arlen, T., Aune, T., et al., 2013, ApJ, 776, 69
\bibitem[Archambault et al. (2014)]{Archambault14} Archambault, S., Aune, T., Behera, B., et al., 2014, ApJ, 785L, 16
\bibitem[Arlen et al. (2013)]{Arlen13} Arlen, T., Aune, T., Beilicke, M., et al., 2013, ApJ, 762, 92

\bibitem[Bartoli et al. (2011)]{Bartoli11} Bartoli, B., Bernardini, P., Bi, X. J., et al., 2011, ApJ, 734, 110
\bibitem[Bartoli et al. (2012)]{Bartoli12} Bartoli, B., Bernardini, P., Bi, X. J., et al., 2012, ApJ, 758, 2
\bibitem[Biteau \& Williams (2015)]{Biteau15} Biteau, J., \& Williams, D. A., 2015, ApJ, 812, 60

\bibitem[Capelo \& Natarajan (2007)]{Capelo07}Capelo, P. R. \& Natarajan, P., 2007, NJPh, 9, 445
\bibitem[Cha et al. (2014)]{Cha14} Cha, Y. J., Zhang, H. J., Zhang, X., et al., 2014, Ap\&SS, 349, 895
\bibitem[Chandra et al. (2010)]{Chandra10} Chandra, P., Yadav, K. K., Rannot, R. C., et al., 2010, JPhG, 37l5201C
\bibitem[Chandra et al. (2012)]{Chandra12} Chandra, P., Rannot, R. C., Yadav, K. K., et al., 2012, JPhG, 39d5201C
\bibitem[Chang et al. (2017)]{Chang17} Chang, Y. L., Arsioli, B., Giommi, P., et al., 2017, A\&A, 598A, 17
\bibitem[Comastri et al. (1997)]{Comastri97} Comastri, A., Fossati, G., Ghisellini, G., \& Molendi, S. 1997, ApJ, 480, 534
\bibitem[Czerny \& Elvis (1987)]{Czerny87} Czerny, B., \& Elvis, M., 1987, ApJ, 321, 305

\bibitem[Daniel et al. (2005)]{Daniel05} Daniel, M. K., Badran, H. M., Bond, I. H., et al., 2005, ApJ, 621, 181
\bibitem[Dermer et al. (1992)]{Dermer92} Dermer C. D., Schlickeiser R. et al., 1992, A\&A, 256, 27
\bibitem[Dermer \& Schlickeiser (1993)]{Dermer93} Dermer C. D. \& Schlickeiser R., 1993, ApJ, 416, 458

\bibitem[Evans et al. (2014)]{Evans14} Evans, P. A., Osborne, J. P., Beardmore, A. P., et al., 2014, ApJS, 210, 8

\bibitem[Fan (2005)]{Fan05} Fan, J.H., 2005, ChJAA (RAA), 5, 213
\bibitem[Fan et al. (2009)]{Fan09} Fan, J. H., Huang, Y., He, T. M., et al., 2009, PASJ, 61, 639
\bibitem[Fan et al. (2012)]{Fan12} Fan, J. H., Yang, J. H., Yuan, Y. H., 2012, ApJ, 761, 125
\bibitem[Fan et al. (2015)]{Fan15} Fan, J. H., Yang, J. H., Liu, Y., et al., 2015, IJMPA, 3045020F
\bibitem[Fan et al. (2016)]{Fan16} Fan, J. H., Yang, J. H., Liu, Y., et al., 2016, ApJS, 226, 20
\bibitem[Fossati et al. (1998)]{Fossati98} Fossati, G., Maraschi, L., Celotti, A., et al., 1998, MNRAS, 299, 433

\bibitem[Giannios et al. (2009)]{Giannios09} Giannios, D., Uzdensky, D. A., Begelman, M. C., 2009, MNRAS, 395L, 29
\bibitem[Giommi et al. (2002)]{Giommi02} Giommi, P., Capalbi, M., Fiocchi, M., et al., 2002, babs.conf, 63
\bibitem[Giommi et al. (2012)]{Giommi12} Giommi, P., Polenta, G., L\"ahteenm\"aki, A., et al., 2012, A\&A, 541A, 160
\bibitem[Godambe et al. (2008)]{Godambe08} Godambe, S. V., Rannot, R. C., Chandra, P., et al., 2008, JPhG, 35f, 5202

\bibitem[Hartman et al. (2001)]{Hartman01} Hartman, R. C., et al., 2001, ApJ, 558, 583
\bibitem[H.E.S.S. Collaboration (2010)]{H.E.S.S.10} H.E.S.S. Collaboration, Abramowski, A., Acero, F., et al., 2010, A\&A, 520A, 83
\bibitem[H.E.S.S. Collaboration (2013)]{H.E.S.S.13} H.E.S.S. Collaboration, Abramowski, A., Acero, F., et al., 2013, MNRAS, 434, 1889
\bibitem[Holder, J. (2012)]{Hold12} Holder, J., 2012, APh, 39, 61
\bibitem[Hovatta et al. (2009)]{Hova09} Hovatta, T., Valtaoja, E., Tornikoski, M., L\"ahteenm\"aki, A., 2009, A\&A, 494, 527
\bibitem[Hovatta et al. (2015)]{Hova15} Hovatta, T., Petropoulou, M., Richards, J. L., 2015, MNRAS, 448, 3121

\bibitem[Jones et al. (1974)]{Jones74} Jones T. W. et al., 1974, ApJ, 188, 353

\bibitem[Kang et al. (2014)]{Kang14} Kang, S. J., Chen, L., Wu, Q. W., 2014, ApJS, 215, 5
\bibitem[Krawczynski et al. (2001)]{Krawczynski01} Krawczynski, H., Sambruna, R., Kohnle, A., 2001, ApJ, 559, 187
\bibitem[Kubo et al. (1998)]{Kubo98} Kubo, H., Takahashi, T., Madejski, G., et al., 1998, ApJ, 504, 693

\bibitem[Landa et al. (1986)]{Landau86} Landau, R., Golisch, B. J., Terry J., 1986, ApJ, 308, 78
\bibitem[Lin \& Fan (2016)]{Lin16} Lin, C. \& Fan, J. H., 2016, RAA, 16, 103
\bibitem[Lin et al. (2017)]{Lin17} Lin, C., Fan, J. H., Xiao H. B., 2017, RAA, 17, 66 (arXiv:170306566L)

\bibitem[Mannheim et al. (1995)]{Mannheim95} Mannheim, K., Schulte, M., \& Rachen, J., 1995, A\&A, 303, L41
\bibitem[Marscher \& Gear (1985)]{Marscher85} Marscher A. P., \& Gear, W. K., 1985, ApJ, 298, 114
\bibitem[Mastichiadis et al. (1997)]{Mastichiadis97} Mastichiadis, A. \& Kirk, J. G., 1997, A\&A, 320, 19
\bibitem[Massaro et al. (2004a)]{Massaro04a} Massaro, E., Perri, M., Giommi, P., 2004a, A\&A, 422, 103
\bibitem[Massaro et al. (2004b)]{Massaro04b} Massaro, E., Perri, M., Giommi, P., \& Nesci, R. 2004b, A\&A, 413, 489
\bibitem[Massaro et al. (2006)]{Massaro06} Massaro, E., Tramacere, A., Perri, M., et al., 2006, A\&A, 448, 861
\bibitem[Massaro et al. (2013)]{Massaro13} Massaro, F., Paggi, A., Errando, M., 2013, ApJS, 207, 16
\bibitem[Meyer et al. (2011)]{Meyer11} Meyer, E. T., Fossati, G., Georganopoulos, M., \& Lister, M. L. 2011, ApJ, 740, 98
\bibitem[M\"ucke et al. (2003)]{M¨¹cke03} M\"ucke, A., Protheroe, R. J., Engel, R., et al., 2003, APh, 18, 593

\bibitem[Nieppola et al. (2006)]{Niep06} Nieppola, E., Tornikoski, M., Valtaoja, E., 2006, A\&A, 445, 441

\bibitem[Padovani \& Giommi (1995)]{Pado92} Padovani, P. \& Giommi, P. 1995, ApJ, 444, 567
\bibitem[Piner et al. (2010)]{Piner10} Piner, B. G., Pant, N., Edwards, P. G., 2010, ApJ, 723, 1150
\bibitem[Pounds et al. (1995)]{Pounds95} Pounds, K. A., Done, C., \& Osborne, J. P. 1995, MNRAS, 277, L5

\bibitem[Rees et al. (1967)]{Ruan67} Rees M. J., 1967, MNRAS, 137, 429
\bibitem[Ross et al. (1992)]{Ross92} Ross, R. R., Fabian, A. C., \& Mineshige, S., 1992, MNRAS, 58, 189

\bibitem[Sahu et al. (2016)]{Sahu16} Sahu, S., Miranda, L. S., Rajpoot, S., 2016, EPJC, 76, 127
\bibitem[Sahu et al. (2016)]{Sahu16} Sambruna, R. M., Maraschi, L., Urry, C. M., 1996, ApJ, 463, 444
\bibitem[Schroedter et al. (2005)]{Schroedter05} Schroedter, M., Badran, H. M., Buckley, J. H., et al., 2005, ApJ, 634, 947
\bibitem[Sharma et al. (2015)]{Sharma15} Sharma, M., Nayak, J., Koul, M. K., et al., 2015, NIMPA, 770, 42
\bibitem[Shields (1978)]{Shields78} Shields, G. A., 1978, Nature, 272, 706
\bibitem[Sikora et al. (1994)]{Sikora94} Sikora, M., Begelman, M. C., \& Rees, M. J. 1994, ApJ, 421, 153
\bibitem[Sikora et al. (2001)]{Sikora01} Sikora, M., et al., 2001, ESASP, 459, 259
\bibitem[Sokolov \& Marscher (2005)]{Sokolov05} Sokolov, A., \& Marscher, A. P., 2005, ApJ, 629, 52

\bibitem[Tramacere et al. (2007)]{Trama07} Tramacere, A., Massaro, F., \& Cavaliere, A. 2007, A\&A, 466, 521
\bibitem[Tramacere et al. (2009)]{Trama09} Tramacere, A., Giommi, P., Perri, M., Verrecchia, F., \& Tosti, G. 2009, A\&A, 501, 879

\bibitem[Urry \& Padovani (1995)]{Urry95} Urry, C. M., Padovani, P., 1995, PASP, 107, 803

\bibitem[Wakely \& Horan (2008)]{Wakely08} Wakely, S. P. \& Horan, D., 2008, ICRC, 3.1341
\bibitem[Wang et al (1996)]{Wang96} Wang, J. C., Luo, Q., \& Xie, G. Z. 1996, ApJ, 457, L65
\bibitem[Weekes, T. C. (1997)]{Weekes97} Weekes, T. C., 1997, AAS, 191, 3701
\bibitem[Wu et al. (2011)]{Wu11} Wu, Q. W., Zou, Y. C., Cao, X. W., 2011, ApJ, 740L, 21
\bibitem[Wu et al. (2009)]{Wu09} Wu, Z. Z., Gu, M. F., \& Jiang, D. R., 2009, RAA, 9, 168

\bibitem[Xiong et al. (2013)]{Xiong13} Xiong, D. R., Zhang, H. J., Zhang, X., et al., 2013, ApSS, 343, 345
\bibitem[Xue et al. (2016)]{Xue16} Xue, R., Luo, D., Du, L. M., et al., 2016, MNRAS, 463, 3038

\bibitem[Yang et al. (2017)]{Yang2017a} Yang, J. H., Fan, J. H., Liu, Y., et al., 2017a, Ap\&SS, 362, 219
\bibitem[Yang et al. (2017)]{Yang2017b} Yang, J. H., Fan, J. H., Liu, Y., et al., 2017b, Ap\&SS, 362, 22

\bibitem[Zhang et al. (2002)]{Zhang02} Zhang, L., Fan, J. H., Cheng, K. S., 2002, PASJ, 54, 159
\bibitem[Zhang et al. (2012)]{Zhang12} Zhang, J., Liang, E. W., Zhang, S. N., Bai, J. M., 2012, ApJ, 752, 157

\end{thebibliography}
\end{document}